# Electromagnetic modes and resonances of two-dimensional bodies

Carlo Forestiere[1], Giovanni Gravina[2], Giovanni Miano[1], Mariano Pascale[1], and Roberto Tricarico[1]
[1]Department of Electrical Engineering and Information Technology
Università degli Studi di Napoli Federico II, via Claudio 21, Napoli, 80125, Italy
[2]Italian Air Force Academy, Via San Gennaro, Agnano 30, 80078 Pozzuoli Napoli

**Abstract**

The electromagnetic modes and the resonances of homogeneous, finite size, two-dimensional bodies are examined in the frequency domain by a rigorous full wave approach based on an integro-differential formulation of the electromagnetic scattering problem. Using a modal expansion for the current density that disentangles the geometric and material properties of the body the integro-differential equation for the induced surface (free or polarization) current density field is solved. The current modes and the corresponding resonant values of the surface conductivity (eigen-conductivities) are evaluated by solving a linear eigenvalue problem with a non-Hermitian operator. They are inherent properties of the body geometry and do not depend on the body material. The material only determines the coefficients of the modal expansion and hence the frequencies at which their amplitudes are maximum (resonance frequencies). The eigen-conductivities and the current modes are studied in detail as the frequency, the shape and the size of the body vary. Open and closed surfaces are considered. The presence of vortex current modes, in addition to the source-sink current modes (no whirling modes), which characterize plasmonic oscillations, is shown. Important topological features of the current modes, such as the number of sources and sinks, the number of vortexes, the direction of the vortexes are preserved as the size of the body and the frequency vary. Unlike the source-sink current modes, in open surfaces the vortex current modes can be resonantly excited only in materials with positive imaginary part of the surface conductivity. Eventually, as examples, the scattering by two-dimensional bodies with either positive or negative imaginary part of the surface conductivity is analyzed and the contributions of the different modes are examined.

## I. Introduction

In the last ten years, the rise of two-dimensional materials has attracted a great amount of interest in the field of plasmonics and photonics (e.g., [1]-[3]). These materials constitute an attractive platform for the engineering of light–matter interactions across the visible, infrared, and terahertz spectral ranges, with new optical control beyond what can be done with bulk materials (e.g., [4]-[7]). In terms of electrical properties, they range from the insulating hexagonal boron nitride and semiconducting transition metal dichalcogenides, to semi metallic graphene. They include multilayers, heterostructures and layered thin films whose thicknesses vary from one atomic layer to tens of nanometers. In general, when the thickness of the body is much smaller than the other two linear dimensions and than the wavelength of the electromagnetic field, only the in-plane electromagnetic response of the material is important and the electromagnetic scattering can be modeled by considering the body as two-dimensional (e.g., [8]).

The understanding of the resonant electromagnetic behavior of a body is a fundamental prerequisite for the engineering of the electromagnetic field-matter interaction. As far as we know this paper studies, for the first time, the modes and the resonances in the electromagnetic scattering from two-dimensional (2D) bodies by using the full Maxwell equations. Modes and resonances are determined by a rigorous full wave approach that naturally discriminates the role of the geometry, of the material, and of the incident electromagnetic field. This is a promising way to design separately the material, the geometry and the incident field for the desired body-electromagnetic field interaction.

When the electromagnetic field is confined within a bounded domain (closed systems) the definition of resonant modes is straightforward. The mathematical model of such problems is characterized by self-adjoint Helmholtz operators leading to a countable infinite set of eigenfunctions, directly related to the resonances of the system. In contrast, when the electromagnetic field occupies the entire space (open systems) the definition of electromagnetic modes and resonance frequencies is more difficult.

The electromagnetic scattering from finite size bodies can be modeled in the frequency domain by means of full wave integral formulations in which the support of problem unknowns is bounded. These formulations, therefore, allow naturally surmounting the difficulties in the analysis of modes and resonances that arise when the spatial domain is unbounded. Different choices of the modes are possible. The quasi-normal modes (e.g., [9]) as well as the characteristic modes [10] are used to study open systems (e.g., [11], [12]). The quasi-normal modes are solution of an intrinsic nonlinear eigenvalue problem. They are not orthogonal in the usual sense; they depend on the material, shape and size of the scattering object. The characteristic modes have been recently evaluated for closed impedance surfaces [13], [14]. They are real and satisfy a weighted orthogonality. The characteristic modes depend on frequency, the shape, size, and material composition of the scattering object. In this paper the concept of material-independent mode [15]-[17] is applied for the first time to the study of the electromagnetic modes and resonances of an isotropic, homogeneous and not space dispersive 2D bodies. The material-independent modes allow separating the role of geometry, material and incident electromagnetic field. This fact provides fundamental information on the resonant electromagnetic behaviors of 2D bodies that other approaches hide.

The surface (free or polarization) current density $\mathbf{j}_s$ induced on the body surface $\Sigma$ is governed, in the frequency domain, by the full wave integro-differential equation (see Section II.1 for more details)

$$\sigma^{-1}\mathbf{j}_s(\mathbf{r}_s) - \zeta_0 \mathcal{L}\{\mathbf{j}_s\}(\mathbf{r}_s) = \mathbf{E}_{inc}^{\parallel}(\mathbf{r}_s) \text{ for any } \mathbf{r}_s \in \Sigma, \qquad (1)$$

where $\sigma$ is the surface conductivity of the body, $\zeta_0$ is the vacuum impedance, $\zeta_0 \mathcal{L}\{\cdot\}$ is the linear integro-differential operator that expresses the tangential component to $\Sigma$ of the induced electric field as function of the induced current density $\mathbf{j}_s$ and $\mathbf{E}_{inc}^{\parallel}$ is the tangential component to $\Sigma$ of the incident electric field. Equation (1) it is nothing but the constitutive relation of the material. It is very important to point out that the linear operator $\mathcal{L}$ does not depend on the material surface conductivity. The operator $\mathcal{L}$ is not Hermitian due to the retardation, hence its eigenfunctions $\{\mathbf{J}_n(\mathbf{r}_s)\}$ are bi-orthogonal and the corresponding eigenvalues $\{1/\sigma_n\}$ are complex (see Section II.2 for more details). The eigenfunctions and the eigenvalues of $\mathcal{L}$ only depend on the body geometry and the frequency.

In this paper equation (1) is solved analytically by using the eigenfunctions $\{\mathbf{J}_n\}$ and the eigenvalues $\{1/\sigma_n\}$ of the linear operator $\mathcal{L}$. The set of eigenfunctions $\{\mathbf{J}_n\}$ are used as basis to represent the surface current density. By exploiting the bi-orthogonality of the set $\{\mathbf{J}_n\}$ the solution of equation (1) is expressed as (more details are given in Section II.3)

$$\mathbf{j}_s(\mathbf{r}_s) = \sum_{n=1}^{\infty} \frac{\sigma_n}{\sigma_n - \zeta_0 \sigma} \langle \mathbf{J}_n^* | \sigma \mathbf{E}_{inc}^{\parallel} \rangle \mathbf{J}_n(\mathbf{r}_s) \qquad (2)$$

where $\langle \cdot | \cdot \rangle$ denotes the scalar product (defined in (19)). This formula disentangles the geometric and material properties of the body and effectively predicts the resonant behavior of 2D bodies as their shape, size and surface conductivity vary. The expansion coefficient of the *n*-th eigenfunction $\mathbf{J}_n$ is proportional to $1/(\sigma_n - \zeta_0 \sigma)$, hence the eigenvalue $\sigma_n$ can be interpreted as the resonant value of the normalized surface conductivity for the mode $\mathbf{J}_n$. For these reasons $\{\sigma_n\}$ are called *eigen-conductivities* of the material independent *current modes* $\{\mathbf{J}_n\}$ of the 2D body. The real part of $\sigma_n$ is always positive and it takes into account the radiation losses of the current mode, instead the imaginary part may be either positive or negative, depending on the mode and the linear dimension of the body normalized to the wavelength.

The mode expansion (2), in the quasi-electrostatic limit, leads to the quasi-electrostatic modes described in [18]-[21], and in the weakly "retarded" limit to the corrections for the quasi-electrostatic modes proposed in [22] and [23]. However, the approaches proposed in [18]-[23] can be only applied to bodies whose linear dimensions are much smaller than the wavelength. Furthermore, these approaches disregard the presence of magnetic modes. Instead, the solution (2) can be applied for any value of the ratio between the linear dimensions of the body and the wavelength and properly takes into account the contribution of the magnetic modes. The presence of vortex current modes in addition to the no whirling modes that characterizes plasmonic oscillations in metals (called source-sink current modes in the paper) is shown. These modes are disregarded in the approaches proposed in [18]-[23]. The source-sink current modes, in the long wavelength limit, tend to the quasi-electrostatic modes, which characterize the plasmonic oscillations in metals. Instead, the vortex current modes, in the long wavelength limit, tend to the quasi-stationary magnetic modes, which cannot be described by the quasi-electrostatic approaches proposed in [18]-[23]. Unlike the source-sink current modes, in open surfaces the vortex current modes can be resonantly excited only in materials with positive imaginary part of the surface conductivity (i.e., dielectric materials). Cylindrical vector beams (e.g., [24]) can couple very efficiently with vortex current modes.

In Section II, the electromagnetic scattering problem is first formulated in the frequency domain by means of the integro-differential equation (1) governing the induced surface current density. Then, the general properties of the spectrum of the linear operator $\mathcal{L}$ are examined and, at the end, the integro-differential equation (1) is solved analytically by using the eigenfunctions (current modes) and eigenvalues (eigen-conductivities) of $\mathcal{L}$. Analytical expressions of the current modes and eigen-conductivities of an infinite plane are given. In Section III, the current modes and the eigen-conductivities of circles, equilateral triangles, rectangles and spherical surfaces are analyzed in detail. The eigenfunctions and the eigenvalues of $\mathcal{L}$ are evaluated numerically by applying a finite element method (details are reported in the Supplementary Materials). In Section IV eventually, the resonance behaviors of the scattering from 2D materials with either positive or negative imaginary part of the surface conductivity are examined. The solution (2) is validated by means of full wave numerical codes by comparing the scattering cross sections of a graphene disk and of a silicon disk.

## II. Electromagnetic scattering

A thin body illuminated by a time harmonic electric field $\text{Re}\left[\mathbf{E}_{inc}(\mathbf{r})e^{i\omega t}\right]$ is considered ($\mathbf{r}$ is the position vector of a generic point of space with the origin at a given point). The body material is linear, isotropic, homogeneous and not space dispersive. The thickness $\Delta$ of the body is smaller than its other two linear characteristic dimensions and than the wavelength of the electromagnetic field. In this limit only the in-plane electromagnetic response of the material is important and the body may be treated as it is two-dimensional (e.g., [8]). A 2D material may be characterized in the frequency domain either by the optical dielectric constant $\varepsilon(\omega)$ or, equivalently, by the optical surface conductivity

$$\sigma = i\omega(\varepsilon - \varepsilon_0)\Delta. \tag{3}$$

In this paper the surface conductivity is used. A simple analytical representation of the surface conductivity is

$$\sigma(\omega) = \frac{1}{\pi}\frac{S_f}{i\omega + \tau_f^{-1}} + \frac{1}{\pi}\frac{S_b}{i(\omega - \omega_b^2/\omega) + \tau_b^{-1}} \tag{4}$$

where the first (Drude) and second (Lorentz) terms represent, respectively, the contribution of the free and bound (polarization) charges. The surface conductivity of different 2D materials can be modeled with a suitable choice of the parameters: the weights $S_f$ and $S_b$, the exciton/phonon frequency $\omega_b$ and the phenomenological relaxation times $\tau_f$ and $\tau_b$ (see in references [6] and [7]). The sign of the imaginary part of the surface conductivity depends on the material and on the frequency, while the real part is always positive for passive materials. For conducting materials the imaginary part of $\sigma$ is always negative in the frequency ranges where the effects of interband transitions are negligible. For dielectric materials, the imaginary part of $\sigma$ may be either positive or negative depending on the frequency. For examples, in suspended high doped (gated) single layer graphene (with chemical potentials of the order of hundred of meV ) it results $R_0\sigma \approx -5i$ in the far and mid infrared ranges where $R_0 = \pi\hbar/e^2 \cong 19.9$ k$\Omega$ (e.g., [26]), while in single layer transition metal dichalcogenides it results $R_0\sigma \approx +5i$ in the near infrared range (e.g., [26], [27]).

### II.1 Mathematical model

The surface of the 2D body is denoted by $\Sigma$, $\mathbf{r}_s$ is the position vector of a generic point of $\Sigma$ and $\hat{\mathbf{n}}$ is the normal to $\Sigma$. The electromagnetic scattering by the 2D body is modeled by the integro-differential equation for the induced surface current density $\mathbf{j}_s(\mathbf{r}_s)$. The constitutive relation of the 2D material is

$$\mathbf{j}_s = \sigma \overleftrightarrow{\mathrm{T}} \mathbf{E} \tag{5}$$

where $\overleftrightarrow{\mathrm{T}}$ is the projector that extracts the tangential component of the electric field $\mathbf{E}(\mathbf{r})$ to the oriented surface $\Sigma$,

$$\overleftrightarrow{\mathrm{T}} \mathbf{E} = -\hat{\mathbf{n}} \times \left[ \hat{\mathbf{n}} \times \mathbf{E}(\mathbf{r}_s) \right]. \tag{6}$$

On $\Sigma$ a surface charge arises whose density amplitude $\rho_s$ is given by

$$\rho_s = -\frac{1}{i\omega} \nabla_s \cdot \mathbf{j}_s \tag{7}$$

where $\nabla_s \cdot$ denotes the surface divergence operator on $\Sigma$. The electric field is expressed as $\mathbf{E} = \mathbf{E}_{inc} + \mathbf{E}_{scat}$ where $\mathbf{E}_{scat}(\mathbf{r})$ is the amplitude of the electric field scattered by the object.

Equation (5) expresses $\mathbf{j}_s$ in terms of the electric field $\mathbf{E}$ through the surface conductivity of the material. On the other hand, the scattered electric and magnetic ($\mathbf{H}_{scat}(\mathbf{r})$) fields can be expressed in terms of $\mathbf{j}_s$ by the electromagnetic potentials in the Lorenz gauge,

$$\mathbf{E}_{scat}(\mathbf{r}) = -\frac{1}{\varepsilon_0} \nabla \mathcal{G}\{\rho_s\}(\mathbf{r}) - i\omega\mu_0 \mathcal{G}\{\mathbf{j}_s\}(\mathbf{r}), \tag{8}$$

$$\mathbf{H}_{scat}(\mathbf{r}) = \nabla \times \mathcal{G}\{\mathbf{j}_s\}(\mathbf{r}), \tag{9}$$

where $\varepsilon_0$ is the vacuum permittivity, $\mu_0$ is the vacuum permeability, $\mathcal{G}\{f\}(\mathbf{r})$ is the integral operator

$$\mathcal{G}\{f\}(\mathbf{r}) = \iint_\Sigma g(\mathbf{r} - \mathbf{r}'_s) f(\mathbf{r}'_s) dS', \tag{10}$$

$g(\mathbf{r})$ is the retarded Green function in vacuum

$$g(\mathbf{r}) = \frac{1}{4\pi r} e^{-ikr}, \tag{11}$$

$r = |\mathbf{r}|$, $k = \omega/c_0$ and $c_0$ is the vacuum light velocity. The first term on the right hand side of equation (8) is the contribution of the electric scalar potential to the electric field, while the second term is the contribution of the magnetic vector potential. By combining (5) and (8), the integro-differential equation (1) that governs the induced surface current density field is obtained, where $\mathcal{L}\{\cdot\}$ is the linear integro-differential operator

$$\mathcal{L}\{\mathbf{j}_s\}(\mathbf{r}_s) = \frac{1}{ik} \overleftrightarrow{\mathrm{T}} \left( \nabla \mathcal{G}\{\nabla_s \cdot \mathbf{j}_s\} + k^2 \mathcal{G}\{\mathbf{j}_s\} \right), \tag{12}$$

$\mathbf{E}_{inc}^{\parallel}(\mathbf{r}_s)$ is the projection of the incident electric field onto the surface $\Sigma$,

$$\mathbf{E}_{inc}^{\parallel}(\mathbf{r}_s) = \overleftrightarrow{\mathrm{T}} \mathbf{E}_{inc}, \tag{13}$$

and $\zeta_0$ is the impedance of vacuum. The integro-differential equation (1) is solved with the boundary condition

$$\mathbf{j}_s \cdot \hat{\mathbf{m}} \big|_{\partial \Sigma} = 0, \tag{14}$$

where $\partial \Sigma$ is the contour of $\Sigma$ and $\hat{\mathbf{m}}$ is the outgoing normal to $\partial \Sigma$ belonging to $\Sigma$. This condition arises from the requirement that the average energy of the scattered electromagnetic field must be limited. Once the induced current density has been evaluated by solving equation (1) with the boundary condition (14), the expressions (8) and (9) allow to evaluate the scattered electromagnetic field everywhere.

### II.2 Auxiliary eigenvalue problem

Since the spectrum of the operator $\mathcal{L}$ is countable infinite, the integro-differential equation (1) can be reduced to an algebraic form by using, as basis functions for representing the surface current density, the eigenfunctions of $\mathcal{L}$ that satisfy the boundary conditions (14). Therefore, the linear eigenvalue problem

$$\mathcal{L}\{\mathbf{J}_n\} = \frac{1}{\sigma_n} \mathbf{J}_n, \tag{15}$$

$$\mathbf{J}_n \cdot \hat{\mathbf{m}} \big|_{\partial \Sigma} = 0, \tag{16}$$

is introduced, where $\sigma_n$ is the "eigenvalue" associated to the eigenfunction $\mathbf{J}_n$ and $n = 1, 2, ...$ (for a reason that will be highlighted in the following, the reciprocals of the eigenvalues have been considered). The eigenvalues $\{\sigma_n\}$ and eigenfunctions $\{\mathbf{J}_n\}$ are dimensionless quantities. In particular, for a given surface $\Sigma$, they only depend on its characteristic linear dimension $l_c$, i.e., the radius of the smallest sphere enclosing the surface $\Sigma$, and on the wavenumber $k$ through the normalized size of the body $kl_c = 2\pi(l_c/\lambda)$; $\lambda$ is the vacuum wavelength.

The eigenvalues and the eigenfunctions of the operator $\mathcal{L}$ are characterized by the following important properties.

I. The eigenvalue $\sigma_n$ is related to the electromagnetic field $(\mathbf{\mathcal{E}}_n, \mathbf{\mathcal{H}}_n)$ generated by the current density field $\mathbf{J}_n$. Indeed, it results that

$$\frac{\mathrm{Re}\{\sigma_n\}}{|\sigma_n|^2} = -\frac{2S_n}{\langle \mathbf{J}_n | \mathbf{J}_n \rangle} \leq 0, \tag{17}$$

$$\frac{\mathrm{Im}\{\sigma_n\}}{|\sigma_n|^2} = \frac{4\omega}{\langle \mathbf{J}_n | \mathbf{J}_n \rangle}\left(W_n^{(mag)} - W_n^{(ele)}\right) \tag{18}$$

where

$$\langle \mathbf{W} | \mathbf{U} \rangle = \iint_\Sigma \mathbf{W}^*(\mathbf{r}_s) \cdot \mathbf{U}(\mathbf{r}_s) dS, \tag{19}$$

$W_n^{(mag)}$ and $W_n^{(ele)}$ are, respectively, the averaged magnetic and electric energies associated to $(\mathbf{\mathcal{E}}_n, \mathbf{\mathcal{H}}_n)$,

$$W_n^{(ele)} = \frac{\varepsilon_0}{4} \iiint_{\mathbb{R}^3} |\mathbf{\mathcal{E}}_n|^2 dV, \tag{20}$$

$$W_n^{(mag)} = \frac{\mu_0}{4} \iiint_{\mathbb{R}^3} |\mathbf{\mathcal{H}}_n|^2 dV, \tag{21}$$

and $S_n$ is the averaged power radiated toward the infinite,

$$S_n = \frac{1}{2\zeta_0} \oiint_{S_\infty} |\mathbf{\mathcal{E}}_n|^2 dS < 0. \tag{22}$$

The electric field $\mathbf{\mathcal{E}}_n(\mathbf{r})$ and the magnetic field $\mathbf{\mathcal{H}}_n(\mathbf{r})$ can be evaluated by the expressions (8) and (9), respectively. Since $S_n$, $W_n^{(mag)}$ and $W_n^{(ele)}$ are positive quantities, it results $\mathrm{Re}\{\sigma_n\} \leq 0$, while the imaginary part of $\sigma_n$ has not a definite sign: when $W_n^{(ele)} > W_n^{(mag)}$ the imaginary part of $\sigma_n$ is negative, otherwise it is positive. In particular, $\mathrm{Re}\{\sigma_n\}/|\sigma_n|^2$ is proportional to the power radiated to infinity from the corresponding current distribution. It accounts for the radiative

losses.

II. In the long wavelength limit, i.e., $kl_c \to 0$, for the <u>non-solenoidal</u> eigenfunctions (namely, $\nabla_s \cdot \mathbf{J}_n \neq 0$) it results $\mathcal{L}\{\mathbf{j}_s\} = \mathcal{E}^{(0)}\{\mathbf{j}_s\}$ where

$$\mathcal{E}^{(0)}\{\mathbf{j}_s\} = \frac{1}{ik}\overset{\leftrightarrow}{\mathrm{T}}\left(\nabla \mathcal{U}\{\nabla_s \cdot \mathbf{j}_s\}\right), \tag{23}$$

$$\mathcal{U}\{f\} = \iint_\Sigma g^{(0)}(\mathbf{r} - \mathbf{r}'_s) f(\mathbf{r}'_s) dS', \tag{24}$$

and

$$g^{(0)}(\mathbf{r}) = \frac{1}{4\pi r} \tag{25}$$

is the static Green function in vacuum. In this case, the effects of the magnetic field are negligible; hence the electric field has an electrostatic character. The linear operator $\mathcal{E}^{(0)}$ is skew-Hermitian ($i\mathcal{E}^{(0)}$ is Hermitian), thus its eigenvalues are purely imaginary and the eigenfunctions associated to distinct eigenvalues are orthogonal with respect to the scalar product (19). Therefore, for $kl_c \to 0$ the eigenvalues of the non-solenoidal eigenfunctions of the operator $\mathcal{L}$ are given by

$$\sigma_n(kl_c) = -\frac{kl_c}{\alpha_n} i \tag{26}$$

where $\alpha_n$ are the eigenvalues of the linear and Hermitian operator $l_c\overset{\leftrightarrow}{\mathrm{T}}\left(-\nabla \mathcal{U}\{\nabla_s \cdot \mathbf{j}_s\}\right)$, which are <u>real</u>, <u>positive</u> and <u>independent</u> of $k$ and $l_c$. This is the *electro-quasi static limit* in which the eigenfunctions of $\mathcal{L}$ are both <u>non-solenoidal</u> and <u>irrotational</u> and the corresponding eigenvalues are purely imaginary with negative imaginary part. The electro-quasi-static limit characterizes the plasmon oscillations in conducting materials.

III. In the long wavelength limit for the <u>solenoidal</u> eigenfunctions (namely, $\nabla_s \cdot \mathbf{J}_n = 0$) it results $\mathcal{L}\{\mathbf{j}_s\} = \mathcal{M}^{(0)}\{\mathbf{j}_s\}$, where

$$\mathcal{M}^{(0)}\{\mathbf{j}_s\}(\mathbf{r}_s) = -ik\mathcal{U}\{\mathbf{j}_s\}. \tag{27}$$

The linear operator $\mathcal{M}^{(0)}$ is skew-Hermitian as $\mathcal{E}^{(0)}$. Therefore, for $kl_c \to 0$, the eigenvalues of the solenoidal eigenfunctions are given by

$$\sigma_n(kl_c) = \frac{1}{\beta_n kl_c} i \tag{28}$$

where $\beta_n$ are the eigenvalues of the linear and Hermitian operator $l_c^{-1}\mathcal{U}\{\mathbf{j}_s\}$, which are <u>real</u>, <u>positive</u> and <u>independent</u> of $k$ and $l_c$. This is the *magneto-quasi-static limit* in which the eigenfunctions are both <u>solenoidal</u> and <u>rotational</u>.

IV. Due to the retardation effects, the linear operator $\mathcal{L}$ is not anti Hermitian. As consequence its eigenvalues are complex with real parts different from zero and the eigenfunctions associated to distinct eigenvalues are no longer orthogonal, that is, $\langle \mathbf{J}_m | \mathbf{J}_n \rangle \neq 0$ for $m \neq n$. Nevertheless, the symmetry of the operator $\mathcal{L}$ implies that (the eigenfunctions are normalized)

$$\langle \mathbf{J}_m^* | \mathbf{J}_n \rangle = \delta_{mn}. \tag{29}$$

V. Let us now consider an infinite plane surface $\Sigma$. In this limit, the spectrum of $\mathcal{L}$ becomes continuous, and its eigenvalues and eigenfunctions can be evaluated analytically by using the 2D Fourier transform and the convolution theorem. The eigenfunctions are

$$\mathbf{J}_\mathbf{q}(\mathbf{r}_s) = \mathbf{U}_\mathbf{q} \exp(-i\mathbf{q} \cdot \mathbf{r}_s) \text{ for any } \mathbf{q} \in \mathbb{R}^2, \tag{30}$$

where $\mathbf{U}_\mathbf{q}$ is solution of the equation

$$\frac{1}{ik}\left[(k^2 - \mathbf{qq})\tilde{g}(\mathbf{q},k)\right]\mathbf{U}_\mathbf{q} = \frac{1}{\sigma_\mathbf{q}}\mathbf{U}_\mathbf{q}, \tag{31}$$

$$\tilde{g}(\mathbf{q},k) = \frac{1}{2\sqrt{q^2 - k^2}} \tag{32}$$

is the 2D Fourier transform of $g(\mathbf{r}_s)$, and $q = |\mathbf{q}|$. For $q \geq k$ the function $\tilde{g}$ is real and it describes evanescent waves in the orthogonal direction to the surface $\Sigma$ with decay length $1/\sqrt{q^2 - k^2}$. While for $q \leq k$ the function $\tilde{g}$ is imaginary, with negative imaginary part, and it describes waves radiated toward infinite with wavenumber $\sqrt{k^2 - q^2}$, which remembers the well known leaky waves (e.g., [28]).

The vector $\mathbf{U}_\mathbf{q}$ is decomposed in the component $\mathbf{U}_\mathbf{q}^\parallel$, parallel to $\mathbf{q}$, and the component $\mathbf{U}_\mathbf{q}^\perp$, orthogonal to $\mathbf{q}$. Both $\mathbf{U}_\mathbf{q}^\parallel$ and $\mathbf{U}_\mathbf{q}^\perp$ are solutions of equation (31) whose eigenvalues are, respectively,

$$\sigma_\mathbf{q}^\parallel = -\frac{2ik}{\sqrt{q^2 - k^2}} \tag{33}$$

and

$$\sigma_\mathbf{q}^\perp = -\frac{2\sqrt{q^2 - k^2}}{ik}. \tag{34}$$

The eigenfunctions $\mathbf{J}_\mathbf{q}^\parallel$ are curl free, while the eigenfunctions $\mathbf{J}_\mathbf{q}^\perp$ are divergence free. In the long wavelength limit $k/q \to 0$, it results $\sigma_\mathbf{q}^\parallel = -i(2k/q)$ and $\sigma_\mathbf{q}^\perp = i(2q/k)$ in agreement with equations (26) and (28), respectively. Furthermore, in the short wavelength limit $k/q \to \infty$, it results $\sigma_\mathbf{q}^\parallel = \sigma_\mathbf{q}^\perp = -2$.

The physical and mathematical meaning of eigenvalues and eigenfunctions of $\mathcal{L}$ is the following. The operator $\zeta_0 \mathcal{L}$ expresses the tangential component of the scattered electric field to $\Sigma$ in terms of the induced surface current density. Consequently, the eigenfunction $\mathbf{J}_n$ is the distribution of the surface current density that coincides with the tangential component to the surface $\Sigma$ of the electric field it generates, apart from the multiplicative factor $1/\sigma_n$. Therefore, the eigenvalue $\sigma_n$ is the value that $\zeta_0 \sigma$ should have so that the corresponding eigenfunction $\mathbf{J}_n$ is a source free solution of equation (1). Since the eigenvalue $\sigma_n$ is complex and has negative real part because of the radiation losses, an active medium is required to excite the *current mode* $\mathbf{J}_n$. In spite of that, the mathematical importance of the current modes $\{\mathbf{J}_n\}$ stems from the fact that they are a basis to represent the solution of equation (1) with the boundary condition (14). In particular, the property (29) is very important because it allows to solve equation (1) in a closed form. The surface current density is given by expression (2). The expansion (2) reveals the important physical mechanisms involved in the electromagnetic scattering and can greatly improve the way it is understood and optimized. The surface conductivity enters through the algebraic fractions $\sigma_n \sigma / (\sigma_n - \zeta_0 \sigma)$. For passive materials, i.e., $\text{Re}\{\sigma\} > 0$, the denominators in (2) cannot vanish because $\text{Re}\{\sigma_n\} < 0$ and the maximum of the current mode amplitude is finite. By using active materials, $|\sigma_n(\omega l_c / c_0) - \zeta_0 \sigma(\omega)|$ could be reduced, in principle, at will. For assigned geometry and surface conductivity $\sigma(\omega)$ it is clear that, if $\langle \mathbf{J}_n^* | \mathbf{E}_{inc}^\parallel \rangle \neq 0$, the amplitude of the current mode $\mathbf{J}_n$ increases as $\zeta_0 \sigma$ approaches $\sigma_n$. In this view, denoted as "frequency picture", the resonance frequency is the value of the frequency for which

$$\left| \sigma_n(\omega l_c / c_0) - \zeta_0 \sigma(\omega) \right| = \min_\omega . \quad (35)$$

This picture requires that both $\langle \mathbf{J}_n^* | \mathbf{E}_{inc}^\parallel \rangle \neq 0$ and $\mathbf{J}_n$ vary very slowly on the frequency in the neighbors of the resonance frequencies. This occurs for the investigated scenarios.

It is possible to introduce a complementary view, denoted as "material picture", where the geometry and the operating frequency are assigned and $\Gamma = \zeta_0 \sigma$ is considered as an independent parameter. The resonant conductivities are defined as the values of $\Gamma$ for which

$$\left| \sigma_n(\omega l_c / c_0) - \Gamma \right| = \min_\Gamma . \quad (36)$$

The material picture is particularly relevant because the conductivity of some 2D materials can be tuned chemically, electrostatically and by applying a magnetic field.

In the paper, for these reasons, the eigenvalue $\sigma_n$ is called *resonant surface conductivity* of the current mode $\mathbf{J}_n$ normalized to the vacuum characteristic admittance $1/\zeta_0$ or briefly it is called *eigen-conductivity*. The actual resonant frequency of the current mode $\mathbf{J}_n$ is obtained by matching its resonant surface conductivity with the surface conductivity of the material according to equation (35).

Once $\mathbf{j}_s$ has been computed by using (2), the scattered electric and magnetic fields can be evaluated by expressions (8) and (9), respectively.

### III. Analysis of current modes, eigen-conductivities, resonances and scattering

The electromagnetic field scattered by a body depends on its shape, size, and material, as well as, on the incident electromagnetic field. The shape and the normalized size of the body determine the current modes and the corresponding eigen-conductivities. The material of the body determines the amplitudes of the current modes for a given incident field according to expression (2). The current mode $\mathbf{J}_n$ is resonantly excited if the frequency satisfies equation (35), provided that $\langle \mathbf{J}_n^* | \mathbf{E}_{inc}^\parallel \rangle \neq 0$. The material losses and the radiation losses determine the amplitude and the width of the resonance peaks.

In this Section, the current modes and the corresponding eigen-conductivities of some canonical surfaces either open (circles, triangles, rectangles) or closed (spherical surfaces) are analyzed in detail. They are applied to the study of the resonances in the electromagnetic scattering from 2D materials with either positive or negative imaginary part of the surface conductivity and the contributions of the different current modes are examined. Due to some different behaviors the paper distinguishes between open and closed surfaces.

In the long wavelength limit, i.e., $kl_c \to 0$, the eigen-conductivities are purely imaginary and the current modes split in two subsets: i) the current modes that are at the same time non-solenoidal and irrotational, whose eigen-conductivities have negative imaginary part; ii) the current modes that are at the same time solenoidal and rotational, whose eigen-conductivities have positive imaginary part. Sources and sinks characterize the field lines of the non-solenoidal and irrotational current modes, whereas loops characterize the field lines of the solenoidal and rotational current modes. Retardation effects give arise to non-solenoidal components that transform loops in vortexes. For these reasons, in the paper this classification is adopted. The current modes that in the long wavelength limit tend to non-solenoidal and irrotational fields are called *source-sink current modes* and the current modes that in the long wavelength limit tend to solenoidal and rotational fields are called *vortex current modes*. For the investigated scenarios each source-sink current mode preserves the number of sources and sinks as the normalized size of the body varies and each vortex current mode preserves the number of vortexes together with their directions.

The eigen-conductivities and the current modes have been evaluated by applying the Galerkin method and the Rao-Wilton-Glisson (RWG) functions have been used as basis functions to represent the surface current density (details are reported in the Supplementary Materials). These basis functions guarantee the continuity of the normal component of the surface current density to any edge of the finite element mesh. In the following by using the analytical solution for the plane surface (see Section II.2) the behavior of the eigen-conductivities for $kl_c \to 0$ and $kl_c \to \infty$ are validated. In Section IV, by using full wave numerical codes the solution (2) is validated.

### III.1 Current modes and eigen-conductivities of open surfaces

The imaginary part of the eigen-conductivities of the vortex current modes in open surfaces is always positive, whereas for the source-sink current modes it may be either negative or positive depending on the value of the size parameter $x$. In the following, the principal features of the source-sink and vortex current modes and of the corresponding eigen-conductivities are analyzed for some open surfaces: circles, equilateral triangles and rectangles with different aspect ratios. We consider, besides the current modes and the corresponding eigen-conductivities, also the amplitude of the dipole moment of the current modes,

$$P_n = \frac{1}{l}\left|\iint_\Sigma \mathbf{J}_n dS\right|, \qquad (37)$$

and the surface average of $|\nabla_s \cdot \mathbf{J}_n|$,

$$D_n = \frac{1}{\Sigma}\iint_\Sigma |\nabla_s \cdot \mathbf{J}_n| dS. \qquad (38)$$

The length $l$ is defined in the following. The quantity $P_n$ gives a measure of the electric dipole moment associated to the current mode $\mathbf{J}_n$, while the quantity $D_n$ gives a measure of the average charge density. For these reasons, in the following we call $P_n$ amplitude of the "electric dipole moment" and $D_n$ average "charge density" associated to the current mode $\mathbf{J}_n$.

A) *Source-sink current modes*

The imaginary part of the eigen-conductivities of the source-sink current modes is always negative in the long wavelength limit. In this limit the source-sink current modes accounts for the plasmonic oscillations in 2D conducting materials. The eigen-conductivities of the source-sink current modes and the mode themselves are labeled in such a way that in the long wavelength limit it results $\text{Im}\{\sigma_1\} < \text{Im}\{\sigma_2\} < ... < 0$.

The source-sink current modes and the corresponding eigen-conductivities of the circle and of the equilateral triangle are first analyzed. The radius of the circle and the length of the side of the triangle are indicated with the same symbol $l$. In the Supplementary Materials, the source-sink current modes and the eigen-conductivities of a rectangle with different values of the aspect ratio are examined in detail. Here only the main conclusions are given.

Figures 1 show the behavior of the first four eigen-conductivities of the circle as function of the size parameter $x = 2\pi(l/\lambda)$: the graphs are the loci that each eigen-conductivity spans in the complex plane as $x$ varies. Figures 2 show the same loci for the triangle.

Figures 3-5 show the real part of the source-sink current modes 1&1', 2&2', 3&3' and 4 of the circle at $l/\lambda = 10^{-2}$, $l/\lambda = 1$ and $l/\lambda = 5$. Figures 6-8 show the source-sink current modes 1&1', 2, 3&3' and 4&4' of the equilateral triangle for the same value of $l/\lambda$. The current modes

on the same columns have the same eigen-conductivities, thus they are degenerate. Tables I and II give the values of the eigen-conductivities of these modes. As expected, at $l/\lambda = 10^{-2}$ the real part of the eigen-conductivities is negligible compared with the corresponding imaginary part. In the "retarded" cases $l/\lambda = 1$ and $l/\lambda = 5$, instead, the real part of the eigen-conductivities is of the same order of magnitude or larger than the imaginary part. Tables III and IV give the amplitudes of the electric dipole moment and the average charge densities associated to the first four source-sink current modes. The modes are normalized according to equation (29).

The loci of the eigen-conductivities are open spirals that start at the point $(0,0)$ of the complex plane for $x = 0$ and move to the third quarter as $x$ increases according to the general properties (26) and (17): both the real and the imaginary parts of the eigen-conductivities become negative and decrease. In particular, the eigen-conductivities tend to zero for $x \to 0$ according to $-ix/\alpha_n$ (equation (19)). Tables V and VI give the values of the parameters $\alpha_n$ for the circle and for the equilateral triangle. They only depend on the shape of the body (they do not depend on the size). At certain values of $x$, depending on the specific current mode, the imaginary part of the eigen-conductivity starts to increase and later it changes its sign. The loci cross the real axis, go into the second quarter and end up at the point $(-2,0)$ of the complex plane. As for the infinite plane discussed in Section 2.4, the real part of the eigen-conductivities approaches $-2$ for $x \to \infty$ and the imaginary part tend to zero but from the positive side. The real part of the eigen-conductivities of the source-sink modes is almost equal or less than $-2$ when the imaginary part is positive. In these circumstances, the radiation losses of the source-sink current modes are very important.

For the circle at $l/\lambda = 10^{-2}$ (Figures 3) the source-sink current modes 1&1' have a strong electric dipolar character, the modes 2&2' have a quadrupolar character and the modes 3&3' have a sextupolar character. These modes are degenerate because of the circular symmetry. The field lines of the mode 4 (the first non degenerate mode) are radial. The number of sources and sinks of these modes are conserved as $l/\lambda$ increases (see Figures 3-5) and the field patterns are almost preserved. In particular, as $l/\lambda$ increases the field lines of these modes penetrate toward the center of the circle.

The amplitude of the electric dipole moment of the source-sink modes 1&1' is large, and it is almost constant as the size parameter $x$ varies (Table III). The amplitude of the electric dipole moment of the other three modes is negligible at $l/\lambda = 10^{-2}$ and increases as the size parameter $x$ increases. In particular, at $l/\lambda = 5$ the amplitude of the electric dipole moment of the modes 2&2' become one order of magnitude less than the amplitude of the electric dipole moment of the modes 1&1'. Moreover, the absolute value of the real part of the eigen-conductivity can assume very large values for the modes 2&2', which have high radiation losses.

For the triangle at $l/\lambda = 10^{-2}$ (Figures 6) the source-sink current modes 1&1 have also a strong electric dipolar character, while the mode 2 has a tripolar character. The current modes 3&3' and of the modes 4&4' have a strong dipolar component and look like the fields of higher order electric dipoles. The modes 1&1', 3&3' and 4&4' are degenerate because of the symmetry of the equilateral triangle. Sources and sinks of the field lines appear at the corners of the triangle.

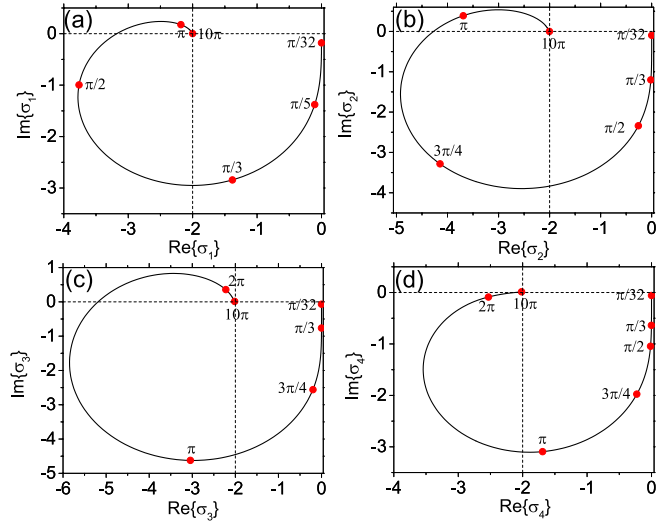

**Figure 1**. Loci of the eigen-conductivities $\sigma_1$ (a), $\sigma_2$ (b), $\sigma_3$ (c) and $\sigma_4$ (d) of the first four source-sink current modes of the circle spanned on the complex plane as function of $x = 2\pi(l/\lambda)$; here $l$ is the circle radius and $\lambda$ is the wavelength in vacuum.

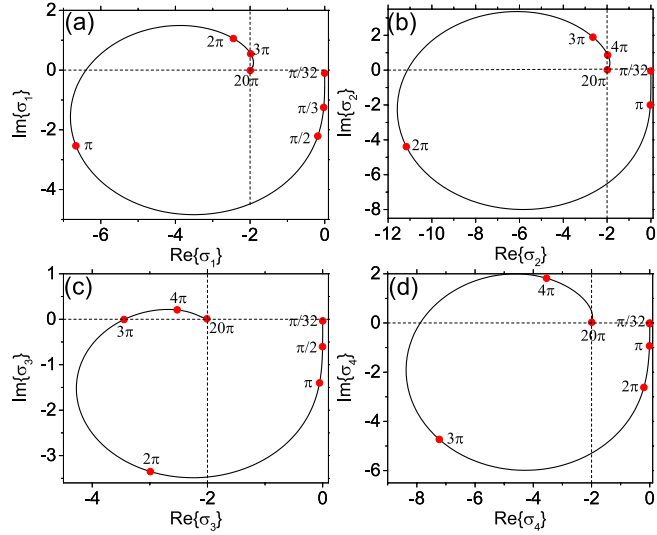

**Figure 2.** Loci of the eigen-conductivities $\sigma_1$ (a), $\sigma_2$ (b), $\sigma_3$ (c) and $\sigma_4$ (d) of the first four source-sink current modes of the triangle spanned on the complex plane as function of $x = 2\pi(l/\lambda)$; here $l$ is the length of the triangle side and $\lambda$ is the wavelength in vacuum.

|      | $l/\lambda = 10^{-2}$ | $l/\lambda = 1$ | $l/\lambda = 5$ |
|------|---|---|---|
| 1&1' | $-7.73 \cdot 10^{-6} - 1.14 \cdot 10^{-1} i$ | $-2.01 + 2.5 \cdot 10^{-2} i$ | $-2.00 + 1.39 \cdot 10^{-2} i$ |
| 2&2' | $-6.50 \cdot 10^{-10} - 6.22 \cdot 10^{-2} i$ | $-2.06 + 1.1 \cdot 10^{-1} i$ | $-2.00 + 1.48 \cdot 10^{-2} i$ |
| 3&3' | $-7.41 \cdot 10^{-14} - 4.33 \cdot 10^{-2} i$ | $-2.22 + 3.5 \cdot 10^{-1} i$ | $-2.00 + 1.60 \cdot 10^{-2} i$ |
| 4    | $-5.55 \cdot 10^{-10} - 3.63 \cdot 10^{-2} i$ | $-2.51 - 9.1 \cdot 10^{-2} i$ | $-2.01 + 1.39 \cdot 10^{-2} i$ |

**Table I**. Eigen-conductivities (the first three digits are given) of the first four source-sink current modes of the circle shown in Figures 3-5; $l$ is the circle radius and $\lambda$ the wavelength in vacuum.

|      | $l/\lambda = 10^{-2}$ | $l/\lambda = 1$ | $l/\lambda = 5$ |
|------|---|---|---|
| 1&1' | $-2.85 \cdot 10^{-7} - 6.79 \cdot 10^{-2} i$ | $-2.43 - 1.06 i$ | $-1.97 + 3.03 \cdot 10^{-2} i$ |
| 2    | $-9.40 \cdot 10^{-12} - 3.20 \cdot 10^{-2} i$ | $-11.6 - 3.89 i$ | $-1.95 + 6.64 \cdot 10^{-2} i$ |
| 3&3' | $-4.79 \cdot 10^{-9} - 2.34 \cdot 10^{-2} i$ | $-3.09 - 3.32 i$ | $-2.04 + 2.11 \cdot 10^{-2} i$ |
| 4&4' | $-2.56 \cdot 10^{-9} - 1.72 \cdot 10^{-2} i$ | $-2.17 \cdot 10^{-1} - 2.68 i$ | $-1.95 + 1.34 \cdot 10^{-1} i$ |

**Table II**. Eigen-conductivities (the first three digits are given) of the first four source-sink current modes of the triangles shown in Figures 6-8; $l$ is the triangle side and $\lambda$ the wavelength in vacuum.

|      | $l/\lambda = 10^{-2}$ | | $l/\lambda = 1$ | | $l/\lambda = 5$ | |
|------|---|---|---|---|---|---|
|      | $P_n$ | $D_n$ | $P_n$ | $D_n$ | $P_n$ | $D_n$ |
| 1&1' | 1.69 | 0.85 | 1.65 | 0.94 | 1.61 | 0.93 |
| 2&2' | $8.4 \cdot 10^{-5}$ | 1.30 | $7 \cdot 10^{-4}$ | 1.44 | $1.47 \cdot 10^{-1}$ | 1.32 |
| 3&3' | $2.4 \cdot 10^{-5}$ | 1.67 | $3 \cdot 10^{-5}$ | 2.17 | $1.53 \cdot 10^{-2}$ | 1.70 |
| 4    | $4.3 \cdot 10^{-7}$ | 2.00 | $1 \cdot 10^{-4}$ | 3.64 | $1.38 \cdot 10^{-2}$ | 1.85 |

**Table III**. Electric dipole moment amplitude $P_n$ and average of the surface charge density amplitude $D_n$ of the source-sink current modes of the circle shown in Figures 3-5; $P_n = \left| \iint_\Sigma \mathbf{J}_n dS \right| / l$, $D_n = \left( \iint_\Sigma \left| \nabla_s \cdot \mathbf{J}_n \right| dS \right) / \Sigma$, $l$ is the circle radius and $\lambda$ the wavelength in vacuum.

|      | $l/\lambda = 10^{-2}$ | | $l/\lambda = 1$ | | $l/\lambda = 5$ | |
|------|---|---|---|---|---|---|
|      | $P_n$ | $D_n$ | $P_n$ | $D_n$ | $P_n$ | $D_n$ |
| 1&1' | $5.4 \cdot 10^{-1}$ | 3.96 | 0.58 | 6.03 | $3.88 \cdot 10^{-1}$ | 10.6 |
| 2    | $1.5 \cdot 10^{-5}$ | 7.84 | $1.4 \cdot 10^{-5}$ | 7.99 | $7.26 \cdot 10^{-4}$ | 13.2 |
| 3&3' | $2.05 \cdot 10^{-1}$ | 9.16 | $2.5 \cdot 10^{-1}$ | 11.2 | $3.58 \cdot 10^{-1}$ | 14.4 |
| 4&4' | $2.02 \cdot 10^{-1}$ | 12.0 | $1.9 \cdot 10^{-1}$ | 12.7 | $2 \cdot 10^{-1}$ | 20.9 |

**Table IV**. Electric dipole moment amplitude $P_n$ and average of the surface charge density amplitude $D_n$ of the source-sink current modes of the triangle shown in Figures 7-8; $P_n = \left| \iint_\Sigma \mathbf{J}_n dS \right| / l$, $D_n = \left( \iint_\Sigma \left| \nabla_s \cdot \mathbf{J}_n \right| dS \right) / \Sigma$, $l$ is the triangle side and $\lambda$ the wavelength in vacuum.

In the modes 2, 3&3' and 4&4', sources and sinks are also located inside the triangle and along the edges. The number of sources and sinks are also conserved as $l/\lambda$ increases (Figures 6-8), but the field patterns change. In particular, the sources and the sinks that in the long wavelength limit are located in correspondence of the corners and edges of the triangle penetrate into it as $l/\lambda$ increases. The amplitude of the electric dipole moment of the modes 1&1', 3&3' and 4&4' is different form zero and vary weakly as $l/\lambda$ increases. The amplitude of the electric dipole moment of the mode 2 is always negligible. Moreover, the absolute value of the real part of the eigen-conductivity may assume very large values for the modes 2&2', which have high radiation losses.

Until now only surfaces whose two characteristic linear dimensions are equal (the circle and the equilateral triangle) have been examined. Now a rectangle is considered and the main difference with respect to a square is briefly highlighted (a detailed description is given in the Supplementary Materials). The behaviors of the source-sink current modes and the eigen-conductivities of a square (namely, a rectangle with aspect ratio equal to one) are very similar to those of a circle. In particular, the first two current modes have a strong electric dipolar character and are degenerate: their electric dipole moments are directed along the two sides of the square. In a rectangle this degeneracy disappears. The first two degenerate modes of the square split into two non-degenerate modes with a strong dipolar character: a *low energy mode* whose electric dipole moment is parallel to the long side of the rectangle and a *high energy mode* whose electric dipole moment is parallel to the short side. High order electric dipolar modes with dipole moments parallel to the long side are between them in terms of energy and their number increases as the aspect ratio of the rectangle increases. They feature current fields oscillating along the long side with either even or odd parity. In the limit of very high aspect ratio the source-sink current modes of the rectangle resemble the modes of a quasi-1D structure [29].

| $n$ | $\alpha_n$ | $\beta_n$ |
|---|---|---|
| 1 | 0.553 | 0.0742 |
| 2 | 1.01 | 0.0666 |
| 3 | 1.45 | 0.0624 |
| 4 | 1.73 | 0.0557 |

**Table V**. Parameters $\alpha_n$ and $\beta_n$ for the circle. The parameters $\alpha_n$ enter in the expression (26), which gives the eigen-conductivities of the source-sink current modes in the long wavelength limit. The parameters $\beta_n$ enter in the expression (28), which gives the eigen-conductivities of the vortex current modes in the long wavelength limit.

| $n$ | $\alpha_n$ | $\beta_n$ |
|---|---|---|
| 1 | 0.149 | 0.372 |
| 2 | 0.315 | 0.258 |
| 3 | 0.430 | 0.198 |
| 4 | 0.586 | 0.193 |

**Table VI**. Parameters $\alpha_n$ and $\beta_n$ for the equilateral triangle. The parameters $\alpha_n$ enter in the expression (26), which gives the eigen-conductivities of the source-sink current modes in the long wavelength limit. The parameters $\beta_n$ enter in the expression (28), which gives the eigen-conductivities of the vortex current modes in the long wavelength limit.

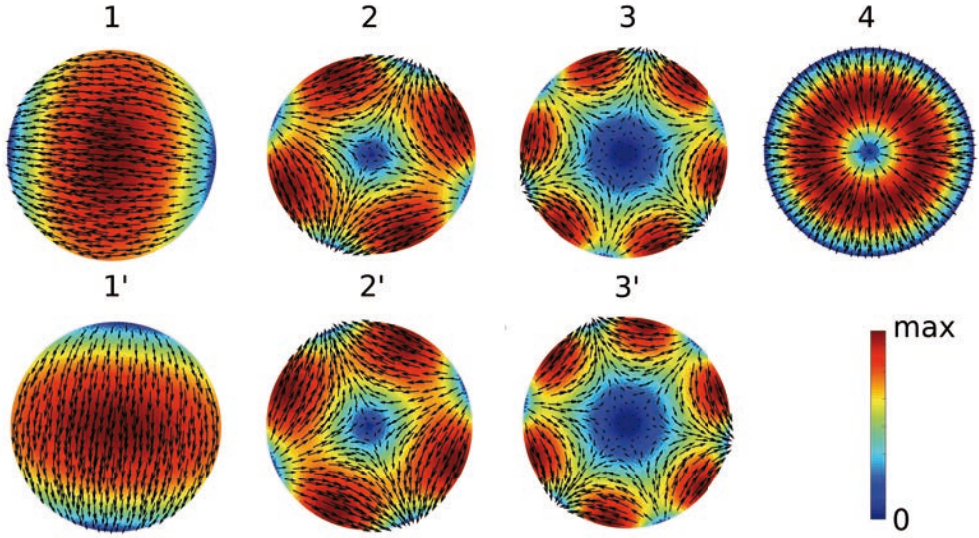

**Figure 3**. Real part of the first four source-sink current modes of the circle at $l/\lambda = 10^{-2}$; here $l$ is the circle radius and $\lambda$ the wavelength in vacuum. The values of corresponding eigen-conductivities are given in Table I, the amplitudes of the electric dipole moments and the averaged surface charge densities are given in Table III. The current modes on the same columns are degenerate. The vectors are directed along $\text{Re}\{\mathbf{J}_n\}$ and the colors represent the intensity of $|\text{Re}\{\mathbf{J}_n\}|$.

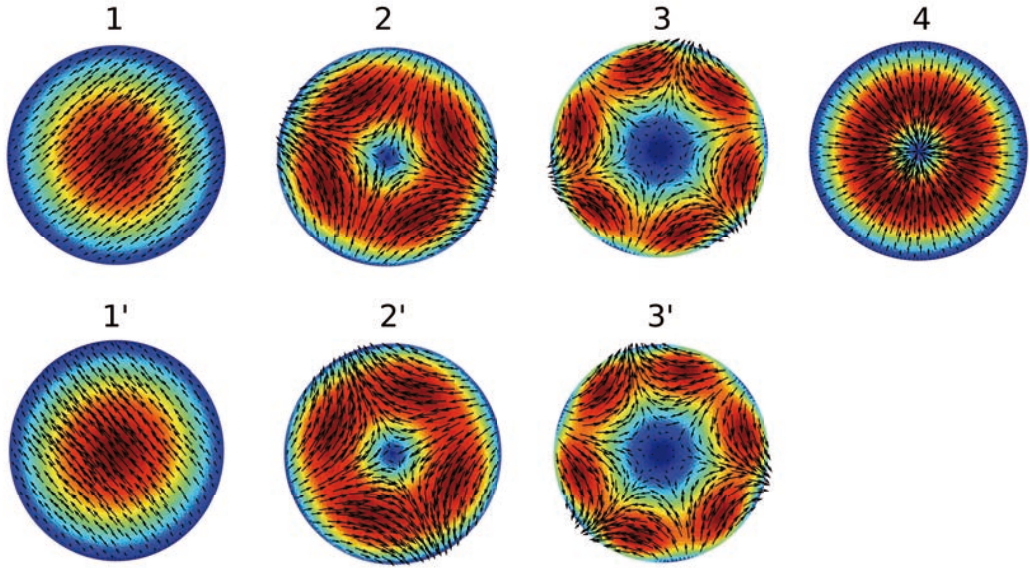

**Figure 4**. Real part of the first four source-sink current modes of a circle at $l/\lambda = 1$; here $l$ is the circle radius and $\lambda$ the wavelength in vacuum. The values of corresponding eigen-conductivities are given in Table I, the amplitudes of the electric dipole moments and the averaged surface charge densities are given in Table III. The current modes on the same columns are degenerate. The vectors are directed along $\text{Re}\{\mathbf{J}_n\}$ and the colors represent the intensity of $|\text{Re}\{\mathbf{J}_n\}|$. The scale of the color intensity is the same shown in Figure 3.

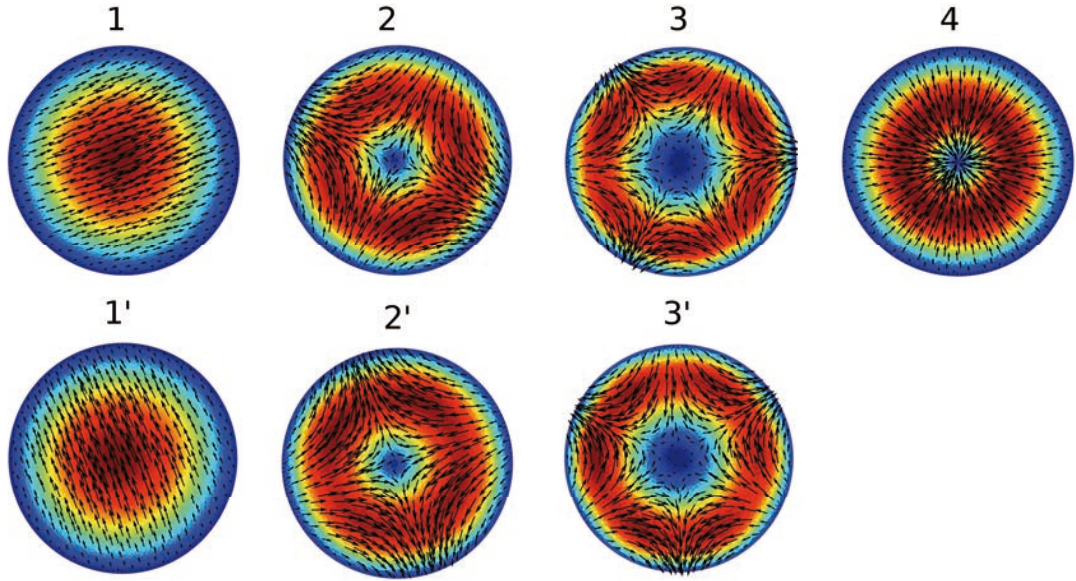

**Figure 5**. Real part of the first four source-sink current modes of a circle at $l/\lambda = 5$; here $l$ is the circle radius and $\lambda$ the wavelength in vacuum. The values of corresponding eigen-conductivities are given in Table I, the amplitudes of the electric dipole moments and the averaged surface charge densities are given in Table III. The current modes on the same columns are degenerate. The vectors are directed along $\text{Re}\{\mathbf{J}_n\}$ and the colors represent the intensity of $|\text{Re}\{\mathbf{J}_n\}|$. The scale of the color intensity is the same shown in Figure 3.

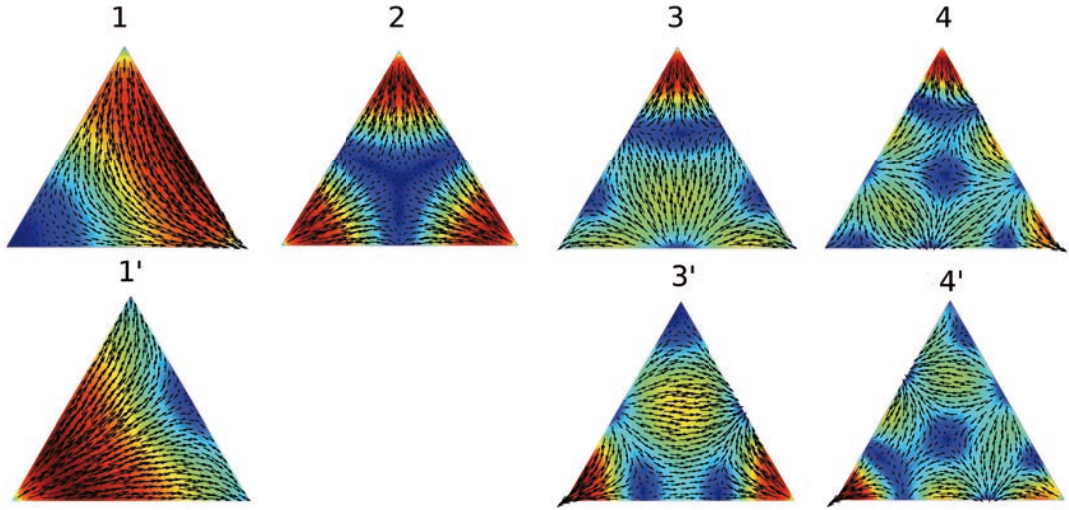

**Figure 6**. Real part of the first four source-sink current modes of a triangle at $l/\lambda = 10^{-2}$; here $l$ is the triangle side and $\lambda$ the wavelength in vacuum. The values of corresponding eigen-conductivities are given in Table II, the amplitudes of the electric dipole moments and the averaged surface charge densities are given in Table IV. The current modes on the same columns are degenerate. The vectors are directed along $\text{Re}\{\mathbf{J}_n\}$ and the colors represent the intensity of $|\text{Re}\{\mathbf{J}_n\}|$. The scale of the color intensity is the same shown in Figure 3.

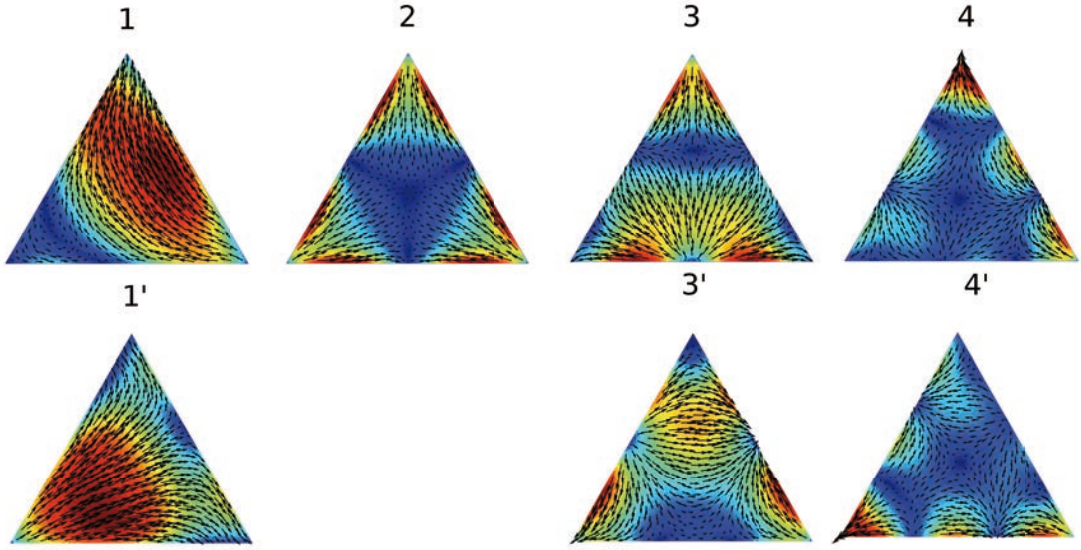

**Figure 7**. Real part of the first four source-sink current modes of a triangle at $l/\lambda = 1$; here $l$ is the triangle side and $\lambda$ the wavelength in vacuum. The values of corresponding eigen-conductivities are given in Table II, the amplitudes of the electric dipole moments and the averaged surface charge densities are given in Table IV. The current modes on the same columns are degenerate. The vectors are directed along $\mathrm{Re}\{\mathbf{J}_n\}$ and the colors represent the intensity of $|\mathrm{Re}\{\mathbf{J}_n\}|$. The scale of the color intensity is the same shown in Figure 3.

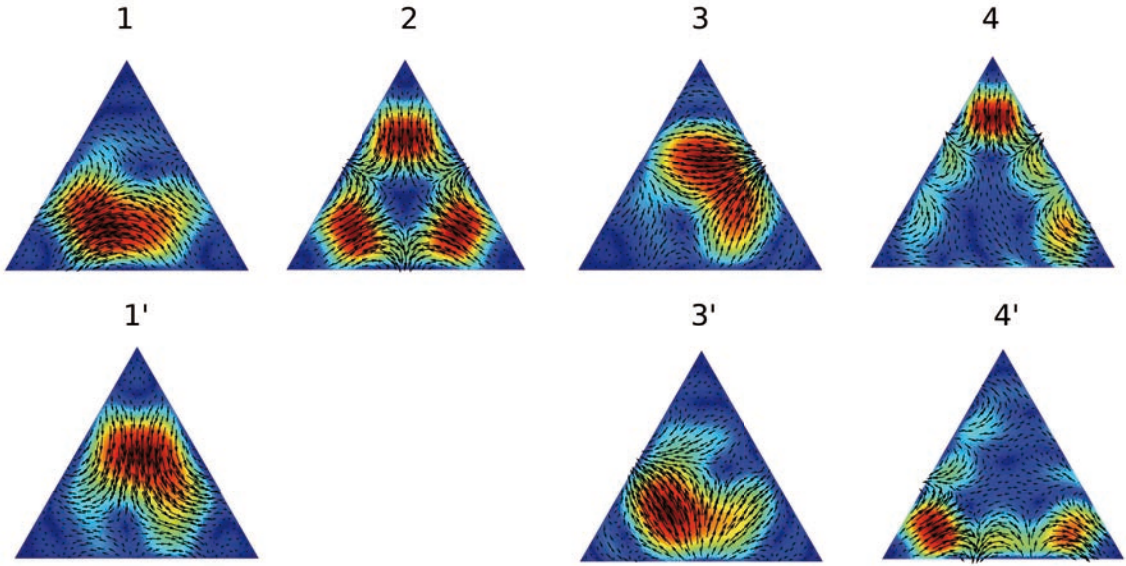

**Figure 8**. Real part of the first four source-sink current modes of a triangle at $l/\lambda = 5$; here $l$ is the triangle side and $\lambda$ the wavelength in vacuum. The values of corresponding eigen-conductivities are given in Table II, the amplitudes of the electric dipole moments and the averaged surface charge densities are given in Table IV. The current modes on the same columns are degenerate. The vectors are directed along $\mathrm{Re}\{\mathbf{J}_n\}$ and the colors represent the intensity of $|\mathrm{Re}\{\mathbf{J}_n\}|$. The scale of the color intensity is the same shown in Figure 3.

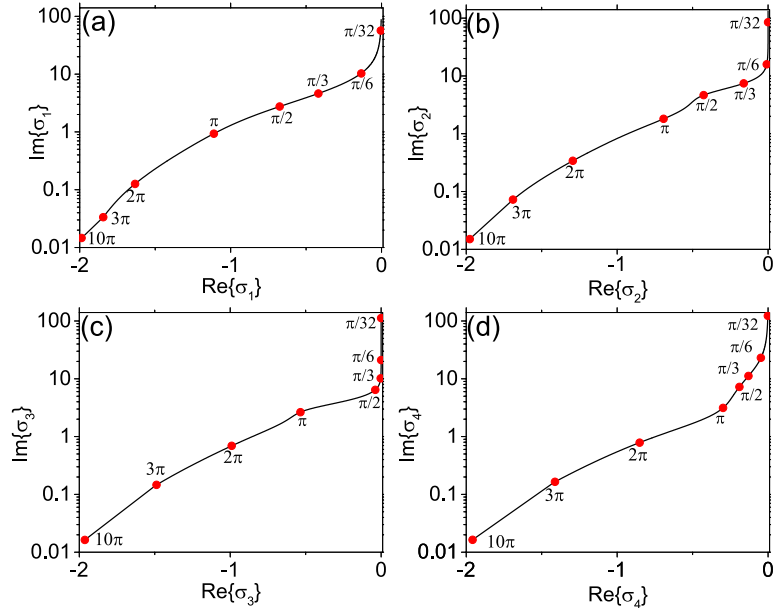

**Figure 9**. Loci of the eigen-conductivities $\sigma_1$ (a), $\sigma_2$ (b), $\sigma_3$ (c) and $\sigma_4$ (d) of the first four vortex current modes of the circle spanned on the complex plane as function of $x = 2\pi(l/\lambda)$; here $l$ is the radius of the circle and $\lambda$ is the wavelength in vacuum.

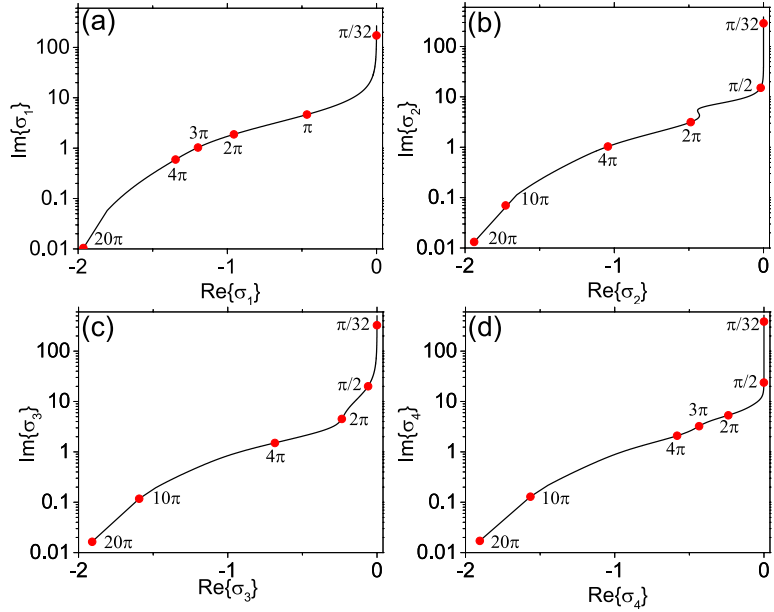

**Figure 10**. Loci of the eigen-conductivities $\sigma_1$ (a), $\sigma_2$ (b), $\sigma_3$ (c) and $\sigma_4$ (d) of the first four vortex current modes of the triangle spanned on the complex plane as function of $x = 2\pi(l/\lambda)$; here $l$ is the length of the triangle side and $\lambda$ is the wavelength in vacuum.

|   | $l/\lambda = 10^{-2}$ | $l/\lambda = 1$ | $l/\lambda = 5$ |
|---|---|---|---|
| 1 | $-2.2 \cdot 10^{-3} + 8.79 \cdot 10^{1} i$ | $-1.63 + 1.24 \cdot 10^{-1} i$ | $-1.99 + 1.46 \cdot 10^{-2} i$ |
| 2&2' | $-1.86 \cdot 10^{-6} + 1.32 \cdot 10^{2} i$ | $-1.30 + 3.36 \cdot 10^{-1} i$ | $-1.98 + 1.51 \cdot 10^{-2} i$ |
| 3&3' | $-1.15 \cdot 10^{-8} + 1.73 \cdot 10^{2} i$ | $-9.95 \cdot 10^{-1} + 6.83 \cdot 10^{-1} i$ | $-1.96 + 1.63 \cdot 10^{-2} i$ |
| 4 | $-8.37 \cdot 10^{-4} + 1.90 \cdot 10^{2} i$ | $-8.56 \cdot 10^{-1} + 7.78 \cdot 10^{-1} i$ | $-1.96 + 1.64 \cdot 10^{-2} i$ |

**Table VII**. Eigen-conductivities (the first three digits are given) of the vortex current modes of the circle shown in Figures 11-13; $l$ is the circle radius and $\lambda$ the wavelength in vacuum.

|   | $l/\lambda = 10^{-2}$ | $l/\lambda = 1$ | $l/\lambda = 5$ |
|---|---|---|---|
| 1 | $-2.69 \cdot 10^{-4} + 2.68 \cdot 10^{2} i$ | $-9.56 \cdot 10^{-1} + 1.87 i$ | $-1.95 + 1.34 \cdot 10^{-1} i$ |
| 2&2' | $-4.42 \cdot 10^{-7} + 3.84 \cdot 10^{2} i$ | $-4.85 \cdot 10^{-1} + 3.12 i$ | $-1.73 + 7 \cdot 10^{-2} i$ |
| 3 | $-1.17 \cdot 10^{-4} + 5.01 \cdot 10^{2} i$ | $-2.33 \cdot 10^{-1} + 4.46 i$ | $-1.59 + 1.17 \cdot 10^{-1} i$ |
| 4&4' | $-1.56 \cdot 10^{-7} + 5.11 \cdot 10^{2} i$ | $-0.32 \cdot 10^{-1} + 4.495 i$ | $-1.56 + 1.29 \cdot 10^{-1} i$ |

**Table VIII**. Eigen-conductivities (the first three digits are given) of the vortex current modes of the triangles shown in Figures 14-16; $l$ is the triangle side and $\lambda$ the wavelength in vacuum.

|   | $l/\lambda = 10^{-2}$ | | $l/\lambda = 1$ | | $l/\lambda = 5$ | |
|---|---|---|---|---|---|---|
|   | $P_n$ | $D_n$ | $P_n$ | $D_n$ | $P_n$ | $D_n$ |
| 1 | $4.2 \cdot 10^{-7}$ | $1.88 \cdot 10^{-6}$ | $4 \cdot 10^{-5}$ | $1.79 \cdot 10^{-2}$ | $1.7 \cdot 10^{-2}$ | $9.05 \cdot 10^{-2}$ |
| 2&2' | $3.5 \cdot 10^{-4}$ | $1.68 \cdot 10^{-4}$ | $5 \cdot 10^{-1}$ | $2.73 \cdot 10^{-1}$ | $3.8 \cdot 10^{-1}$ | $2.51 \cdot 10^{-1}$ |
| 3&3' | $4.4 \cdot 10^{-4}$ | $8.75 \cdot 10^{-5}$ | $1.7 \cdot 10^{-4}$ | $6.11 \cdot 10^{-1}$ | $5 \cdot 10^{-3}$ | $4.15 \cdot 10^{-1}$ |
| 4 | $4.3 \cdot 10^{-7}$ | $1.88 \cdot 10^{-6}$ | $5 \cdot 10^{-5}$ | $4.15 \cdot 10^{-2}$ | $5.2 \cdot 10^{-3}$ | $1.41 \cdot 10^{-1}$ |

**Table IX**. Electric dipole moment amplitude $P_n$ and average of the surface charge density amplitude $D_n$ of the vortex current modes of the circle shown in Figures 11-13; $P_n = \left| \iint_\Sigma \mathbf{J}_n \, dS \right| / l$, $D_n = \left( \iint_\Sigma |\nabla_s \cdot \mathbf{J}_n| \, dS \right) / \Sigma$, $l$ is the circle radius and $\lambda$ the wavelength in vacuum.

|   | $l/\lambda = 10^{-2}$ | | $l/\lambda = 1$ | | $l/\lambda = 5$ | |
|---|---|---|---|---|---|---|
|   | $P_n$ | $D_n$ | $P_n$ | $D_n$ | $P_n$ | $D_n$ |
| 1 | $1 \cdot 10^{-7}$ | $1.15 \cdot 10^{-4}$ | $5.1 \cdot 10^{-6}$ | $1.61$ | $7.1 \cdot 10^{-5}$ | $3.52$ |
| 2&2' | $2 \cdot 10^{-5}$ | $1.81 \cdot 10^{-4}$ | $1.4 \cdot 10^{-1}$ | $1.77$ | $1.07 \cdot 10^{-1}$ | $4.49$ |
| 3&3 | $5.4 \cdot 10^{-8}$ | $3.39 \cdot 10^{-5}$ | $1 \cdot 10^{-4}$ | $0.47$ | $2.07 \cdot 10^{-4}$ | $5.98$ |
| 4 | $7 \cdot 10^{-6}$ | $1.04 \cdot 10^{-4}$ | $7.3 \cdot 10^{-2}$ | $1.65$ | $3.95 \cdot 10^{-2}$ | $6.09$ |

**Table X**. Electric dipole moment amplitude $P_n$ and average of the surface charge density amplitude $D_n$ of the of the vortex current modes of the triangle shown in Figures 14-16; $P_n = \left| \iint_\Sigma \mathbf{J}_n \, dS \right| / l$, $D_n = \left( \iint_\Sigma |\nabla_s \cdot \mathbf{J}_n| \, dS \right) / \Sigma$. $l$ is the triangle side and $\lambda$ the wavelength in vacuum.

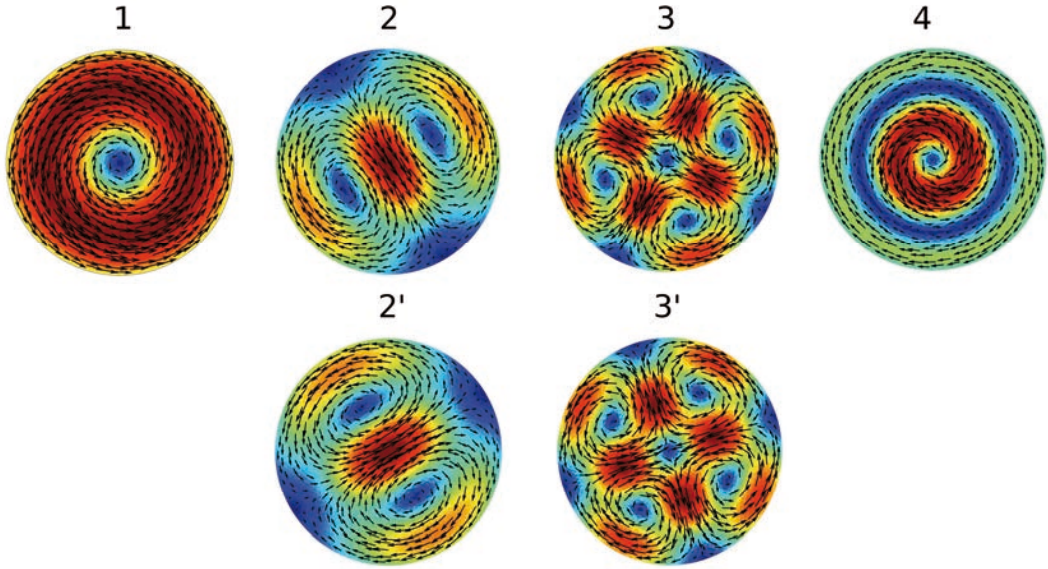

**Figure 11**. Real part of the first four vortex current modes of a circle for $l/\lambda = 10^{-2}$; here $l$ is the circle radius and $\lambda$ the wavelength in vacuum. The values of corresponding eigen-conductivities are given in Table VII, the amplitudes of the electric dipole moments and the averaged surface charge densities are given in Table VIII. The current modes on the same columns are degenerate. The vectors are directed along $\text{Re}\{\mathbf{J}_n\}$ and the colors represent the intensity of $|\text{Re}\{\mathbf{J}_n\}|$. The scale of the color intensity is the same shown in Figure 3.

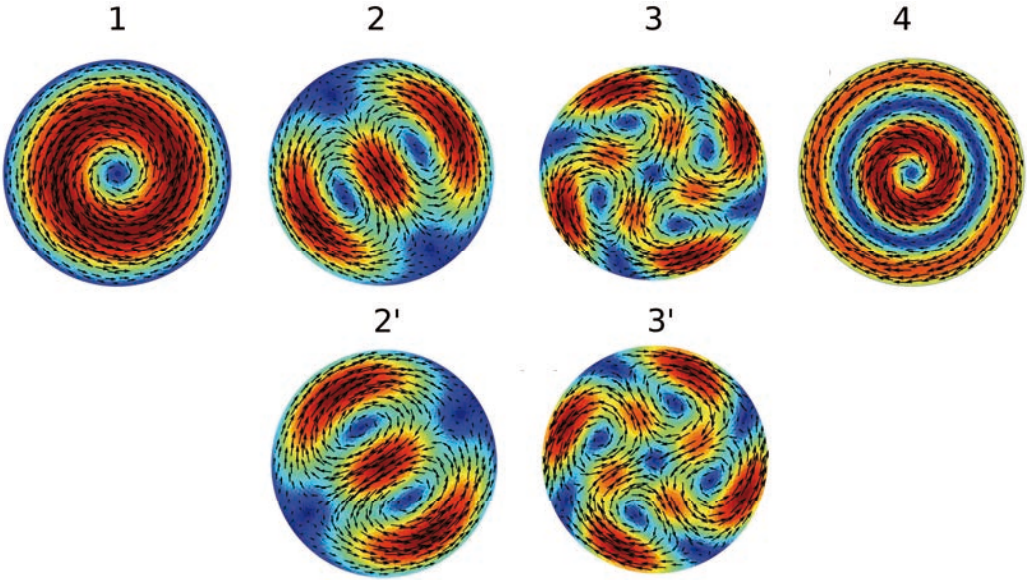

**Figure 12**. Real part of the first four vortex current modes of a circle for $l/\lambda = 1$; here $l$ is the circle radius and $\lambda$ the wavelength in vacuum. The values of corresponding eigen-conductivities are given in Table VII, the amplitudes of the electric dipole moments and the averaged surface charge densities are given in Table VIII. The current modes on the same columns are degenerate. The vectors are directed along $\text{Re}\{\mathbf{J}_n\}$ and the colors represent the intensity of $|\text{Re}\{\mathbf{J}_n\}|$. The scale of the color intensity is the same shown in Figure 3.

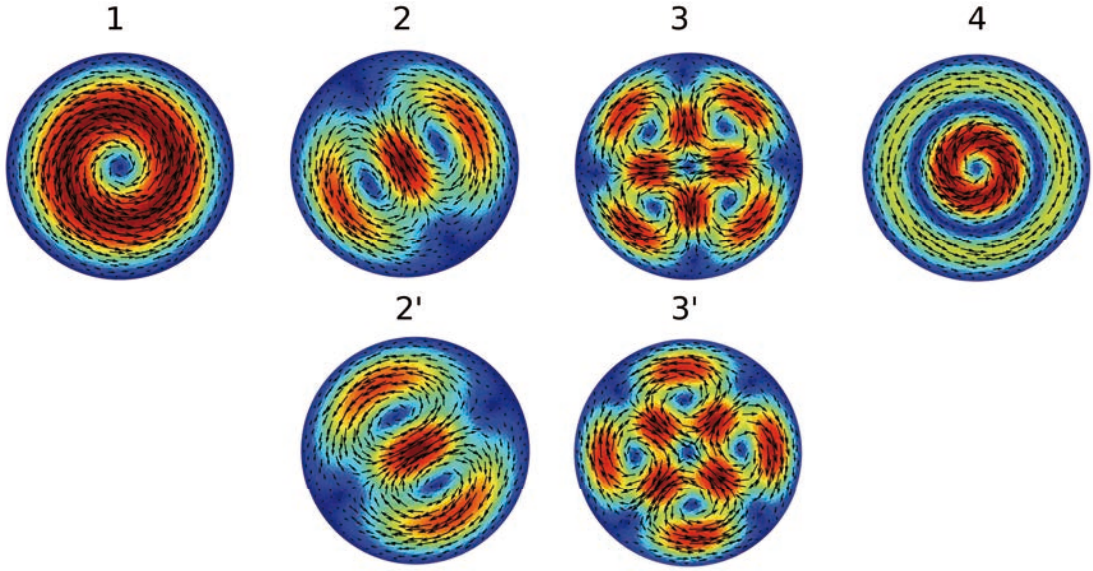

**Figure 13**. Real part of the first four vortex current modes of a circle for $l/\lambda = 5$; here $l$ is the circle radius and $\lambda$ the wavelength in vacuum. The values of corresponding eigen-conductivities are given in Table VII, the amplitudes of the electric dipole moments and the averaged surface charge densities are given in Table VIII. The current modes on the same columns are degenerate. The vectors are directed along $\text{Re}\{\mathbf{J}_n\}$ and the colors represent the intensity of $|\text{Re}\{\mathbf{J}_n\}|$. The scale of the color intensity is the same shown in Figure 3.

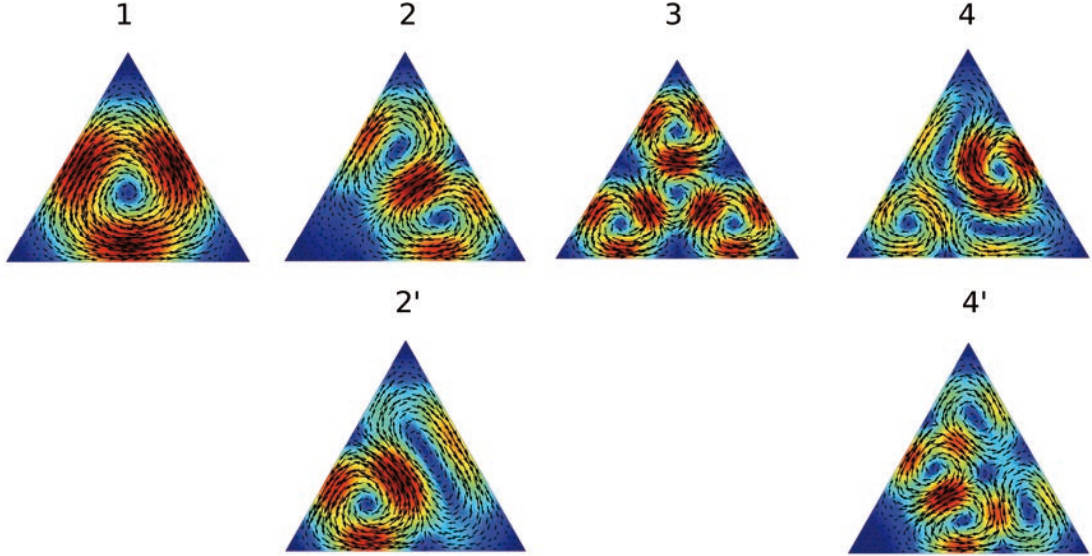

**Figure 14**. Real part of the first four vortex current modes of a triangle for $l/\lambda = 10^{-2}$; here $l$ is the triangle side and $\lambda$ the wavelength in vacuum. The values of corresponding eigen-conductivities are given in Table IX, the amplitudes of the electric dipole moments and the averaged surface charge densities are given in Table X. The current modes on the same columns are degenerate. The vectors are directed along $\text{Re}\{\mathbf{J}_n\}$ and the colors represent the intensity of $|\text{Re}\{\mathbf{J}_n\}|$. The scale of the color intensity is the same shown in Figure 3.

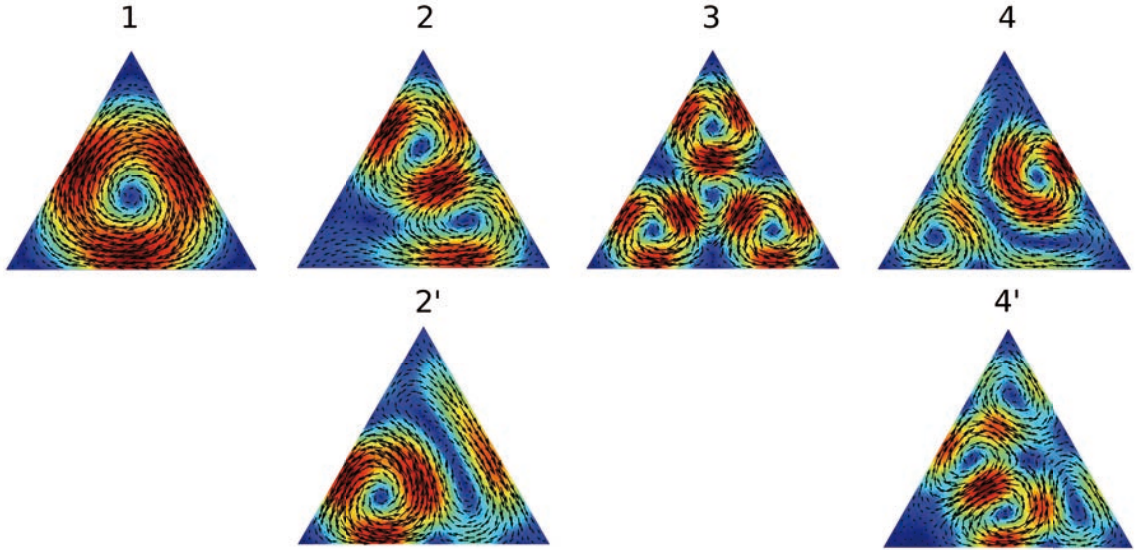

**Figure 15**. Real part of the first four vortex current modes of a triangle for $l/\lambda = 1$; here $l$ is the triangle side and $\lambda$ the wavelength in vacuum. The values of corresponding eigen-conductivities are given in Table IX, the amplitudes of the electric dipole moments and the averaged surface charge densities are given in Table X. The current modes on the same columns are degenerate. The vectors are directed along $\text{Re}\{\mathbf{J}_n\}$ and the colors represent the intensity of $|\text{Re}\{\mathbf{J}_n\}|$. The scale of the color intensity is the same shown in Figure 3.

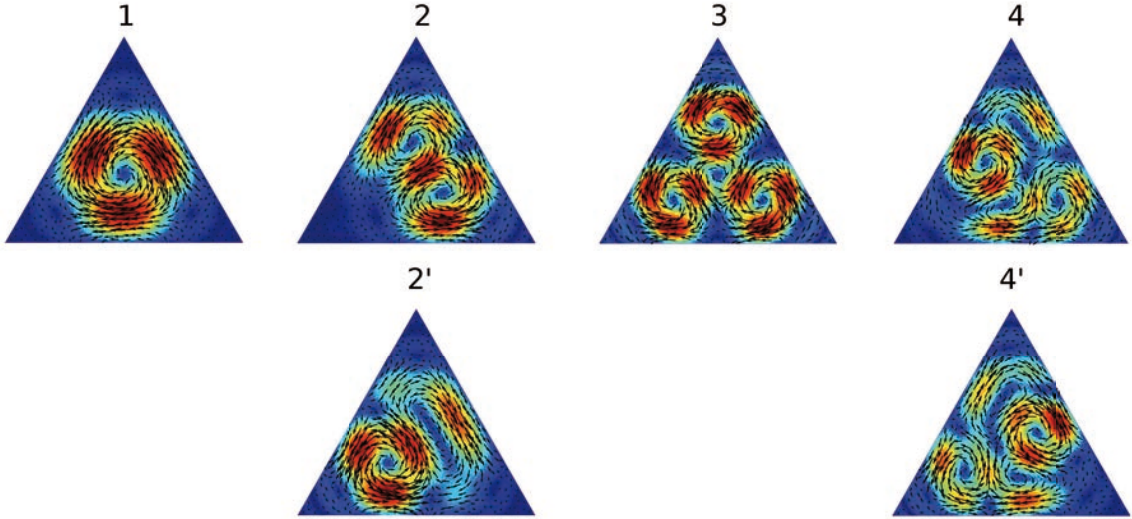

**Figure 16**. Real part of the first four vortex current modes of a triangle for $l/\lambda = 5$; here $l$ is the triangle side and $\lambda$ the wavelength in vacuum. The values of corresponding eigen-conductivities are given in Table IX, the amplitudes of the electric dipole moments and the averaged surface charge densities are given in Table X. The current modes on the same columns are degenerate. The vectors are directed along $\text{Re}\{\mathbf{J}_n\}$ and the colors represent the intensity of $|\text{Re}\{\mathbf{J}_n\}|$. The scale of the color intensity is the same shown in Figure 3.

B) *Vortex current modes*

Vortex current modes are now examined. In the long wavelength limit, the imaginary part of the eigen-conductivities of these modes is positive. Therefore the vortex current mode and the corresponding eigen-conductivities are labeled in such a way that $0 < \text{Im}\{\sigma_1\} < \text{Im}\{\sigma_2\} < ...$ in the long wavelength limit.

The vortex current modes and the corresponding eigen-conductivities of the circle and of the equilateral triangle are first examined. Again the radius of the circle and the length of the side of the triangle are indicated with the same symbol $l$. In the Supplementary Materials, the vortex current modes and the eigen-conductivities of a rectangle with different values of the aspect ratio are examined in detail. Here only the main conclusions are reported.

Figures 11 show the behavior of the eigen-conductivities of the first four vortex current modes for the circle as function of $x = 2\pi(l/\lambda)$ and Figures 12 show the same loci for the triangle. Unlike the loci of the eigen-conductivities of the source-sink current modes, these loci belong always to the second quarter of the complex plane; therefore the imaginary part of the eigen-conductivities is always positive, while the real part is always negative. For $x \to 0$ the eigen-conductivities tend to $+i\infty$ according to $i/(\beta_n x)$ (equation (28)). Tables V and VI give the values of the parameters $\beta_n$ for the circle and for the equilateral triangle. They only depend on the shape of the 2D body (they do not depend on the size). As $x$ increases, both the real and imaginary parts of the eigen-conductivities decrease. The real parts approach $-2$, while the imaginary parts tend to zero. This behavior agrees with that of the infinite plane. By summarizing, the loci of the eigen-conductivities of the vortex current modes start from the point $(0, +\infty)$ of the complex plane for $x = 0$, stay in the second quarter as $x$ increases and end up into the point $(-2, 0)$ for $x \to \infty$. The real part of the eigen-conductivities of the vortex current modes is confined into the interval $(-2, 0)$.

Figures 11-13 show the real part of the vortex current modes 1, 2&2', 3&3', 4 of the circle at $l/\lambda = 10^{-2}$, $l/\lambda = 1$ and $l/\lambda = 5$. Figures 14-16 show the vortex current modes 1, 2&2', 3, 4&4' of the equilateral triangle for the same value of $l/\lambda$. The current modes on the same columns are degenerate. Table VII and Table VIII give, respectively, the values of the eigen-conductivities of these modes. Tables IX and X give the amplitudes of the electric dipole moments and the average of the surface charge density amplitudes (the modes are normalized according to equation (29)).

As for source-sink current modes, at $l/\lambda = 10^{-2}$ the real part of the eigen-conductivities is always negligible compared with the imaginary part, while in the "retarded" cases, $l/\lambda = 1$ and $l/\lambda = 5$, the real part of the eigen-conductivities become comparable or greater than the imaginary part. Moreover, the amplitude of the surface charge density is negligible at $l/\lambda = 10^{-2}$ but it becomes comparable with the amplitude of the surface charge density of the source-sink current modes at $l/\lambda = 1$ and $l/\lambda = 5$. For these reasons the field lines of these modes are not loops but vortexes. At last, the amplitude of the electric dipole moment of the vortex current modes 2&2' both of the circle and of the triangle become large at $l/\lambda = 1$ and $l/\lambda = 5$.

For the circle at $l/\lambda = 10^{-2}$ (Figure 11) the vortex current mode 1 has a strong magnetic dipolar character. Its magnetic dipole moment is oriented orthogonally to the surface. The vortex current modes 2&2' are composed of two identical sets of current vortexes with antiparallel magnetic dipole moments and the third vortex current modes 3&3' are composed of four identical sets of current vortexes with alternating magnetic dipole moments. The current mode 4 consists of concentric current vortexes in which the direction of the current changes at a certain radius. Due to the circular symmetry the modes 2&2' and the modes 3&3' are degenerate. The electric dipole moment of these modes is negligible (Table IX). As $l/\lambda$ increases, the number of vortexes is conserved (Figure 9-11) and the shape of the field lines is qualitatively preserved. As for the source-sink current modes the vortex current modes penetrate into the circle. It is interesting to observe that the amplitude of the electric dipole moment of the vortex current modes 2&2' increases significantly as $l/\lambda$ increases (Table IX). This property allows the excitation of this mode by a plane wave propagating orthogonally to the circle surface.

As for the circle, the vortex current mode 1 of the triangle at $l/\lambda = 10^{-2}$ has a strong magnetic dipolar character (Figure 12). The modes 2&2' and the modes 4&4' are degenerate. The modes 2&2' are composed of two different sets of current loops whose magnetic dipole moments are antiparallel and the modes 3&3' are composed of four identical sets of current loops with alternating magnetic dipole moments. The vortex current modes 4&4' are composed of two different sets of loops and show the same mirror symmetry of the modes 2&2'. As for the circle the number of vortexes is conserved as $l/\lambda$ increases (Figures 12-14), and the mode patterns are substantially preserved. The amplitudes of the electric dipole moments are negligible for the modes 1 and 3, while they reach significant values for the modes 2&2', 4&4' in the "retarded" regime.

Considering a square, we find that the behaviors of the vortex current modes and of the eigen-conductivities are similar to those of the circle. But, if we turn the square into a rectangle, as the ratio between the two edges increases, the vortexes align themselves and two adjacent vortexes are always contra-rotating. A detailed description is given in Supplementary Materials. In the limit of very high aspect ratio the eigen-conductivity of all the vortex modes diverge.

### III.2 Current modes and eigen-conductivities for a closed surface

In this Section, the behavior of the current modes and of the eigen-conductivities for a spherical surface is investigated. In Figure 17, the real parts of the first three source-sink and vortex current modes ($n = 1,2,3$) of a spherical surface are shown. In this case, the distributions of the current modes are independent of the frequency and spherical surface radius. The source-sink modes do not have a radial component of the magnetic field, while the vortex modes do not have a radial component of the electric field. They are therefore transverse magnetic (TM), and transverse electric (TE), respectively.

Now, the behavior of the eigen-conductivities as a function of the size parameter $x = 2\pi(l/\lambda)$ is shown; $l$ is now the radius of the spherical surface. Figures 18(a) and 18(b) show, respectively, the behavior of the real part and imaginary parts of the eigen-conductivities of the first two source-sink current modes as $x$ varies; Figures 18(c) and 18(d) show the same plots for the first two vortex current modes. Similarly to open surfaces, in the long wavelength limit $(l/\lambda \to 0)$ both the source-sink and vortex modes exhibit negligible radiative losses, hence $\text{Re}\{\sigma_n\}$ approaches zero for $x \to 0$. For the source-sink current modes $\text{Re}\{\sigma_n\}$ decreases as $x$ increases and it exhibits a minimum at $x = 1.4$ for $n = 1$ and at $x = 2.4$ for $n = 2$, then it asymptotically approaches the values of $-1$ for $x \to \infty$. On the contrary, for the vortex current modes $\text{Re}\{\sigma_n\}$ monotonically decreases from 0 to $-1$. Once again, this behavior is analogous to the case of an open surface, where the $\mathrm{Re}\{\sigma_n\}$ decreased monotonically for vortex modes, while it had a minimum for source-sink modes. The behavior of $\mathrm{Im}\{\sigma_n\}$ is now described. It is possible to prove that, for a spherical surface, it results:

$$\lim_{x \to 0} \mathrm{Im}\{x^{-1}\sigma_n\} = -\frac{2n+1}{n(n+1)} \quad \text{for source-sink modes,} \tag{39}$$

$$\lim_{x \to 0} \mathrm{Im}\{x\sigma_n\} = 2n+1 \quad \text{for vortice modes.} \tag{40}$$

Therefore, for the sink-current modes $\mathrm{Im}\{\sigma_n\}$ approaches zero for $x \to 0$, while for the vortex current modes $\mathrm{Im}\{\sigma_n\}$ diverge. This is consistent with the results obtained for open surfaces.

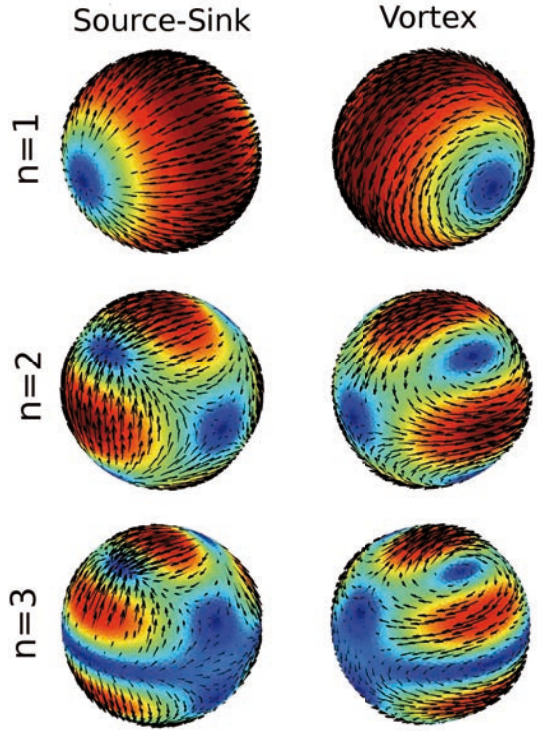

**Figure 17**. Plots of the real part of source-sink and vortex current modes ($n = 1, 2, 3$) of a spherical surface. The direction of each arrow represents the direction of the mode at the corresponding point on the spherical surface, while the length of the arrow and the color describe its amplitude. The scale of the color intensity is the same shown in Figure 3.

In the end, note that $\mathrm{Im}\{\sigma_n\}$ for both source-sink and vortex modes diverges in correspondence of an infinite countable set of the size parameters $x$, which coincide with the resonant values of the normalized radius of a spherical cavity bounded by a perfect electric conductor. Specifically, with black and red dots, Figure 18 highlights the values of $x$ in correspondence of which the cavity resonates. They are given by [30,31]

$$\text{TM}_n \rightarrow \frac{d}{dx}\left[\sqrt{x}J_{n+\frac{1}{2}}(x)\right] = 0, \quad \text{TE}_n \rightarrow J_{n+\frac{1}{2}}(x) = 0, \tag{41}$$

where $J_\nu(x)$ is the Bessel function of first kind. As consequence, for closed surface the limit of $\text{Im}\{\sigma_n(x)\}$ for $x \rightarrow \infty$ does not exist for both the source-sink and vortex current modes; furthermore, note that $\text{Re}\{\sigma_n\} \rightarrow -1$ for $x \rightarrow \infty$. These behaviors differentiate the resonances of closed surfaces from the resonances of open surfaces.

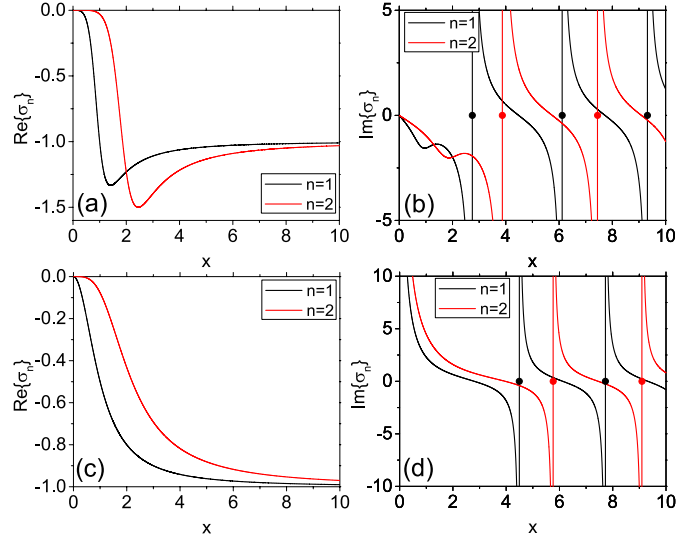

**Figure 18**. Real and imaginary part of the eigen-conductivities $\sigma_1$ and $\sigma_2$ as a function of $x = 2\pi(l/\lambda)$ for a spherical surface with radius $l$ ($\lambda$ is the wavelength in vacuum): source-sink current modes (a-b); vortex current modes (c-d). The black and red dots indicate the values of $x$ in correspondence of which a spherical cavity bounded by a perfect electrical conductor resonates.

## IV. Scattering from 2D object and resonant conditions

The width and the amplitude of the resonance peaks depend on the losses in the material and on the radiation losses. The real part of the eigen-conductivities of both types of current modes becomes negligible only in the long wavelength limit; otherwise, it is in the order of magnitude of 1. Furthermore, for open surface the imaginary part of the eigen-conductivities is always positive for the vortex current modes and it is either positive or negative for the source-sink current modes depending on the values of the normalized body size. In the long wavelength limit it is negative. Therefore, narrow resonant source-sink current modes can be mainly excited in conducting 2D materials and narrow resonant vortex current modes can be only excited in 2D dielectric materials, unless active materials are used.

For the purpose to exemplify, the current mode expansion (2) is now used to study the scattering from a disk illuminated by a linearly polarized plane wave (with unitary amplitude), propagating normally to it: in a graphene disk source-sink current modes may be resonantly excited, while in a silicon thin disk vortex current modes may be excited resonantly. In particular, the scattering efficiency and the amplitude of the near electric field are evaluated and the contributions of the different current modes are highlighted.

### IV.1 Disk with negative imaginary part of the surface conductivity

Firstly an example of material with negative imaginary part of the surface conductivity is considered: a suspended high-doped (or gated) graphene disk with a radius of 100 nm. The electromagnetic field scattered by it in a wavelength range around 10 $\mu$m is analyzed. The incident electromagnetic field is a plane wave linearly polarized, propagating normally to the disk.

For the considered disk radius, the size quantum effects are negligible and, for the considered wavelength range, the interband transitions may be disregarded. Thus the surface conductivity of the graphene is approximated as (e.g., [25])

$$\sigma = \frac{1}{R_0} \frac{\mu}{\hbar(i\omega + 1/\tau)}, \qquad (42)$$

where $R_0 = \pi\hbar/e^2 \cong 19.9$ k$\Omega$, $\mu$ is the chemical potential, and $\tau$ is the electron relaxation time (due to the scattering of electrons with phonons, $\tau \approx 5 \cdot 10^{-13}$ s).

The chemical potential of the graphene is chosen in such a way that the disk resonates at 10 $\mu$m on the first source-sink current mode, which has a strong dipolar character. Since $l/\lambda = 0.01$, the effects of the radiation losses are negligible. Around 10 $\mu$m, the effects of the graphene losses are negligible, too. Therefore, the resonant condition for the $n-th$ source-sink current mode is obtained by choosing the chemical potential in such a way that (material picture):

$$\frac{x}{\alpha_n} \cong \frac{\zeta_0}{R_0} \frac{\mu}{\hbar\omega} \qquad (43)$$

where the values of $\alpha_n$ for the first four source-sink current modes of the disk are given in Table V. For $x = 0.01$ and $\lambda = 10$ $\mu$m this equation gives $\mu \cong 0.492$ eV (at $\lambda = 10$ $\mu$m it results $\zeta_0 \sigma \cong -0.114i$).

Figure 19(a) shows the scattering efficiency. A very good agreement is found between the scattering cross sections obtained by using the analytical solution (2) and the scattering cross section obtained by means of a full wave two-dimensional numerical code [32]. This fact validates the solution (2) and the computation of the current modes and the eigen-conductivities. Besides the designed resonance at 10 $\mu$m, two Fano resonances emerge in both the spectra of the scattering efficiency and the near field amplitude. Figure 19(b) shows the maximum of the amplitude of the electric field on the disk surface. The Figure 19(c) shows the resonant source-sink current modes excited in the graphene disk and the vertical dashed lines in Figures 19(a) and 19(b) indicate their resonant wavelengths evaluated by Equation (35) (frequency picture). The Figure 17(c) shows the electric dipole modes 1&1', the high order electric dipole modes 7&7' and 15&15' (these modes are degenerate). The eigen-conductivities of these modes are $\sigma_1 = -7.80 \cdot 10^{-6} - 0.114i$, $\sigma_7 = -2.6612 \cdot 10^{-7} - 0.053i$, $\sigma_{15} = -4.7314 \cdot 10^{-9} - 0.040i$ and the

corresponding resonance wavelengths are $\lambda_1 = 10\ \mu m$, $\lambda_7 = 4.87\ \mu m$ and $\lambda_{15} = 3.90\ \mu m$. The two Fano resonances are due to the interference between two adjacent electric dipole modes. Top Figure 19(d) shows the distribution of the real part of the electric field on the graphene disk at wavelengths in correspondence of the three peaks of the electric field amplitude. Lower Figure 19(d) shows the corresponding distributions of the induced surface charge density.

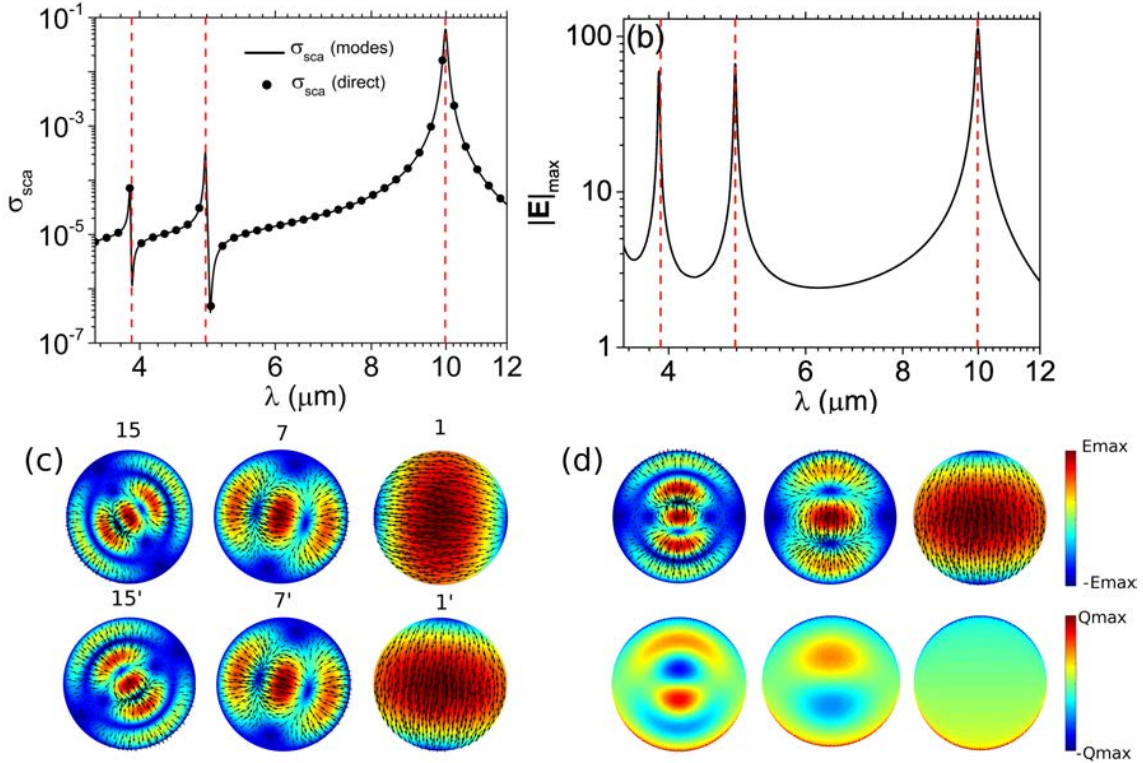

**Figure 19**. (a) Spectra of scattering efficiency and (b) maximum amplitude of electric field on the surface of a graphene disk with radius $l = 100$ nm, excited by a linearly polarized plane wave propagating normally to the disk. The graphene chemical potential has been designed to enforce the resonance of the electric dipolar mode at $\lambda = 10 \mu m$. The vertical dashed lines indicate the resonant wavelength (c) of the source-sink current modes (the scale of the color intensity is the same shown in (d)), calculated by Equation (35) (frequency picture). Top (d) distribution of the electric in correspondence of the three peaks in the near field shown in (b). The vectors are directed along **j** and the colors represent $|\mathbf{j}|$. Below (d) distributions of the corresponding surface charge densities. The black points in the panel (a) indicate the values of the scattering efficiency $\sigma_{sca}^{(direct)}$ obtained by solving directly equation (1) by a full wave 2D numerical code [32].

## IV.2 Disk with positive imaginary part of the surface conductivity

Now an example of material with positive imaginary part of the surface conductivity is considered: a dielectric thin disk with relative permittivity $\varepsilon_r = 16$ (silicon), radius $l = 500$ nm and thickness $\Delta = 19.2$ nm. The electromagnetic field scattered by it in the wavelength range from $0.5\ \mu m$ to $1.25\ \mu m$ is analyzed. The incident electromagnetic field is a plane wave linearly polarized, propagating normally to the disk.

Since the disk thickness $\Delta$ is much smaller than its radius and the wavelength, the thin disk may be represented as a circle with equivalent surface conductivity given by (3),

$$\zeta_0 \sigma = x\left(\frac{\Delta}{l}\right)(\varepsilon_r - 1)i .\qquad(44)$$

Therefore, for the considered dielectric disk, the imaginary part of $\zeta_0 \sigma$ varies in the interval $(1.45, 3.61)$.

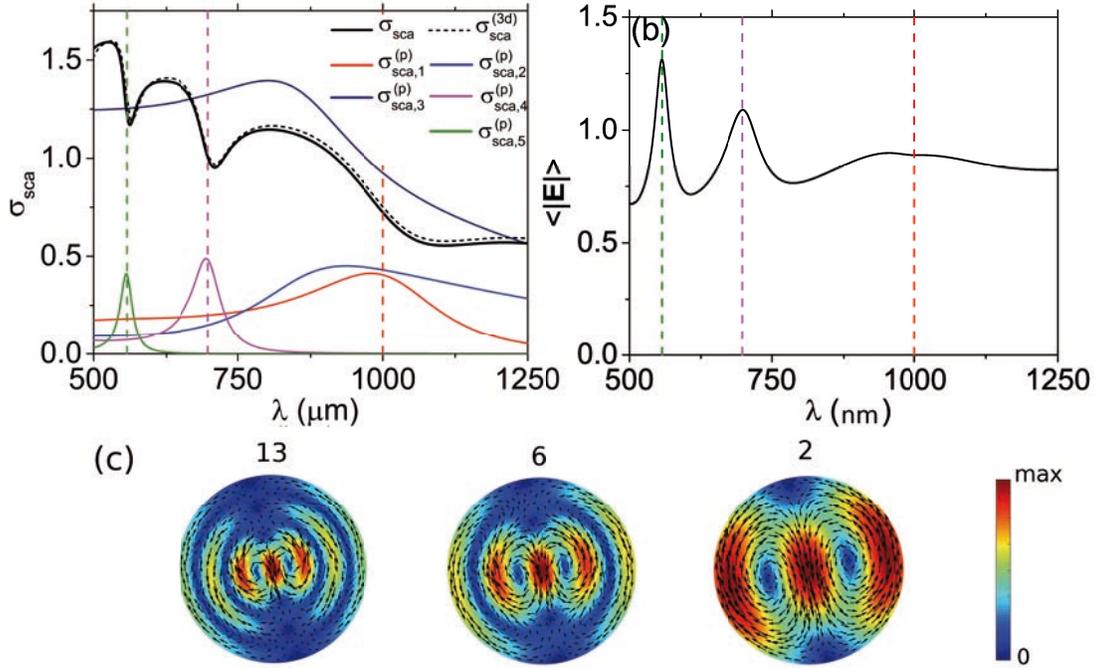

**Figure 20**. Scattering efficiency (a) and averaged near field enhancement (b) of a silicon thin disk with radius $l = 500$ nm and thickness $\Delta = 19.2$ nm, illuminated by a linearly polarized plane wave propagating normally to the disk. The vertical dashed lines indicate the positions of the resonant wavelengths (calculated in the "frequency picture" by Eq. (35)) of the three vortex modes shown in panel (c). The partial scattering cross sections of the most important current modes that contribute to the scattered field are shown: $\sigma_{v,2}^{(p)}, \sigma_{v,6}^{(p)}, \sigma_{v,13}^{(p)}$ are the partial scattering efficiency of the three resonant vortex current modes shown in (c), while $\sigma_{ss,1}^{(p)}, \sigma_{ss,1'}^{(p)}$ are the partial scattering efficiency of the off resonant first source-sink current mode, which is double degenerate. The dashed line in the panel (a) indicates the values of the scattering efficiency $\sigma_{sca}^{(3d)}$ obtained by using a three dimensional full wave numerical code [33].

Figure 20(a) shows the scattering efficiency (black line). Again a very good agreement is found between the scattering cross sections obtained by using the expression (2) and the scattering cross section obtained by means of a full wave three-dimensional numerical code [33]. This fact again validates the solution (2) and the computation of the current modes and the eigen-conductivities. Several current modes contribute to the scattered electromagnetic field. The vertical dashed lines in Figures 20(a) indicate the values of the resonant wavelengths of the vortex current modes shown in Figure 20(c), $\lambda_2 = 980$ nm, $\lambda_6 = 696$ nm and $\lambda_{13} = 556$ nm. Besides these resonant modes, the first two source-sink current modes of the disk (modes 1&1' in Figures 3-5) contribute to the scattered field. These two degenerate modes are off resonance because their eigen-conductivities are bounded in the region of complex plane with the imaginary part belonging to the interval $(-2.95, 0.238)$ regardless of $l/\lambda$, while the imaginary part of $\zeta_0 \sigma$ varies between 1.45 and 3.61. Even if these modes are off resonance, their coupling amplitudes with the plane wave are very high because they have a strong electric dipole moment. To highlight the contribution of the most important current modes, Figure 20(a) also shows the partial scattering efficiencies [34] defined as the scattering efficiency obtained by considering only one mode at a time. The partial scattering efficiencies $\sigma_{v,2}^{(p)}, \sigma_{v,6}^{(p)}, \sigma_{v,13}^{(p)}$ are relevant to the three vortex current modes shown in Figure 20(c); $\sigma_{ss,1}^{(p)}$ and $\sigma_{ss,1'}^{(p)}$ are, instead, the partial scattering efficiency of the first source-sink current modes. The destructive interferences between the broad source-sink modes and one of the narrow vortex current mode shown in Figure 20(c) give rise to the Fano resonances around $\lambda = 1106$ nm, $\lambda = 711$ nm and $\lambda = 563$ nm. Figure 20(b) shows the averaged amplitude of the electric field on the disk surface. The vertical dashed lines always indicate the values of the resonant wavelengths of the vortex current modes shown in Figure 20(c).

## V. Concluding remarks

The electromagnetic modes and resonances in the full wave electromagnetic scattering from 2D bodies have been investigated for the first time. We have used the concept of material-independent modes and it is also the first time that the material picture is applied to study resonances in 2D bodies.

The current modes and the corresponding eigen-conductivities are solution of a linear eigenvalue problem. The eigen-conductivity $\sigma_n$ corresponding to the current mode $\mathbf{J}_n$ is the value that the surface conductivity of the body, normalized to the vacuum admittance $1/\zeta_0$, should have so that the current mode $\mathbf{J}_n$ is a free source solution of the Maxwell equations. The current modes and the corresponding eigen-conductivities only depend on the geometry of the body and on the frequency, they are independent of the surface conductivity $\sigma$. The real part of the eigen-conductivity is proportional to the radiation losses of the current mode; hence, it is always negative. The imaginary part is proportional to the difference between the time average of the magnetic energy and the time average of the electric energy of the mode hence it may be either negative or positive, depending on the shape of the 2D body and on the normalized body size $kl_c$.

The existence of vortex current modes in addition to source-sink current modes (no whirling modes), which describe plasmonic oscillations, is shown ~~demonstrated for the first time~~. The current modes show very important topological features that are preserved as $kl_c$ varies. The source –sink current modes, which in the long wavelength limit $(kl_c \rightarrow 0)$ are irrotational, are characterized by a number of sources and sinks of the field lines that is conserved as the ratio between the characteristic linear dimension of the body and the wavelength $kl_c$ increases. Similarly, the vortex current modes, which in the long wavelength limit are solenoidal, are characterized by a number of vortexes of the field lines that is conserved.

The surface current density induced on the body by an external excitation $\mathbf{E}_{inc}^{\parallel}$ is represented in terms of the current modes. The expansion coefficient of each current mode is equal to $\sigma_n / (\sigma_n - \zeta_0 \sigma) \langle \mathbf{J}_n^* | \sigma \mathbf{E}_{inc}^{\parallel} \rangle$. The resonant frequency of the mode is the frequency for which $|\sigma_n / (\sigma_n - \zeta_0 \sigma)|$ is maximum. This expansion reveals the important physical mechanisms involved in the electromagnetic scattering and can greatly improve the way it is understood and optimized.

Illustrative examples for open surfaces (disk, equilateral triangle, rectangle) and a closed surface (spherical surface) are given. The diagrams on the complex plane of the loci of the eigen-conductivities show some very important general properties. The analysis of these loci is propaedeutic to the understanding of the scattering from the object: they determine the resonance frequencies of the current modes once the material is assigned.

The scattering efficiency and the amplitude of the near electric field of a disk with either positive or negative imaginary part of the surface conductivity were presented and analyzed, and the contribution of the important current modes was highlighted. Unlike the source-sink current modes, in open surfaces the vortex current modes can be resonantly excited only in materials with positive imaginary part of the surface conductivity.

To deal with homogeneous and isotropic materials that are dispersive both in time and space we have to operate in the frequency and wavenumber domains, this is an open problem. Equation (1) can be written also for anisotropic/bianisotropic and/or inhomogeneous 2D bodies, as long as $\sigma^{-1}(\omega)$ is replaced by $\overleftrightarrow{\sigma}^{-1}(\mathbf{r}_s; \omega)$, where $\overleftrightarrow{\sigma}$ is a tensor depending on the position. The shortcoming is that the inhomogeneity and/or the anisotropic/bianisotropic couple the eigenmodes of the linear operator $\mathcal{L}$ and to obtain the solution of equation (1) we have to invert a matrix.

## References

[1] F. Xia, H. Wang, D. Xiao, M. Dubey, A. Ramasubramaniam, Two-dimensional material nanophotonics, *Nature Photonics*, **Vol. 8**, pages 899–907, 2014.
[2] K. S. Novoselov, A. Mishchenko, A. Carvalho, A. H. Castro Neto, 2D materials and van der Waals heterostructures, Science, vol. 353, issue 6298, aac9439, 2016.
[3] Y. Li, Z. Li, C. Chi, H. Shan, L. Zheng and Z. Fang, Plasmonics of 2D nanomaterials: properties and applications, Advanced Science, 4, 1600430, 2017.
[4] Z. Sun, A. Martinez, F. Wang, Optical modulators with 2D layered materials, Nature Photonics, vol. 10, pages 227–238, 2016.
[5] K. F. Mak, J. Shan, Photonics and optoelectronics of 2D semiconductor transition metal dichalcogenides, Nature Photonics, vol.10, pages 216–226, 2016.
[6] D. N. Basov, M. M. Fogler, F. J. García de Abajo, Polaritons in van der Waals materials, Science, Vol. 354, Issue 6309, aag1992, 2016.
[7] T. Low, A. Chaves, J. D. Caldwell, A. Kumar, N. X. Fang, P. Avouris, T. F. Heinz, F. Guinea, L. Martin-Moreno & F. Koppens, Polaritons in layered two-dimensional materials, *Nature Materials,* Vol. 16, pages 182–194, 2017.
[8] R. F. Harrington, J. R. Mautz, An Impedance Sheet Approximation for Thin Dielectric Shells, IEEE Transactions on Antennas and Propagation, Vol. 23 , Issue 4, pp. 531-534, 1975.
[9] P. Lalanne, W. Yan, K. Vynck, C. Sauvan, J.P. Hugonin, Light interaction with photonic and plasmonic resonances, Laser & Photonics Reviews, Vol. 12 , 1700113, 2018.
[10] R. J. Garbacz, Modal expansions for resonance scattering phenomena, Proceedings of the IEEE, vol. 53, pp. 856–864, 1965.
[11] D. A. Powel, Resonant dynamics of arbitrarily shaped meta-atoms, Phys. Rev. B 90, 075108, 2014.

[12] J. Mäkitalo, M. Kauranen, and S. Suuriniemi, Modes and resonances of plasmonic scatters, Phys. Rev. B 89, 165429, 2014.
[13] P. Ylä-Oijala, J. Lappalainen, and S. Järvenpää, Characteristic Mode Equations for Impedance Surfaces, IEEE Transactions on Antennas and Propagation, Vol. 66, pp. 487-492, 2018.
[14] A. Sihvola, D. C. Tzarouchis, P.Ylä-Oijala, H. Wallén, and B. Kong, Resonances in small scatterers with impedance boundary, arXiv:1805.09159v1, 2018.
[15] D. J. Bergman and D. Stroud, Theory of resonances in the electromagnetic scattering by macroscopic bodies, Phys. Rev. B 22, 3527, 1980.
[16] C. Forestiere, G. Miano, Material-independent modes for electromagnetic scattering, Phys. Rev. B 94, 201406, 2016.
[17] C. Forestiere, G. Miano, G. Rubinacci, A. Tamburrino, R. Tricarico, and S. Ventre, Volume Integral Formulation for the Calculation of Material Independent Modes of Dielectric Scatterers, IEEE Trans. Antennas Propagation, Vol. 66, pp. 2505-2514, 2018.
[18] D. R. Fredkin and I. D. Mayergoyz, Resonant Behavior of Dielectric Objects (Electrostatic Resonances), Phys. Rev. Lett. 91, 253902, 2003.
[19] I. Silveiro, J. M. Plaza Ortega and F. J. García de Abajo, Plasmon wave function of graphene nanoribbons, New Journal of Physics, Vol. 17, 083013, 2015.
[20] W. Wang, T. Christensen, A. P. Jauho, K. S. Thygesen, M. Wubs, N. A. Mortensen, Plasmonic eigenmodes in individual and bow-tie graphene nanotriangles, Scientific Reports, 5, 9535, 2015.
[21] R. Yu, J. D. Cox, J. R. M. Saavedra, J. Garcia de Abajo, Analytical Modeling of Graphene Plasmons, ACS Photonics, Vol. 4, pp. 3106-3114, 2017.
[22] L. Hung, S. Y. Lee, O. McGovern, O. Rabin, and I. Mayergoyz, Calculation and measurement of radiation corrections for plasmon resonances in nanoparticles, Phys. Rev. B 88, 075424, 2013.
[23] R. Yu, L. M. Liz-Marzánbcd, F. J. García de Abajo, Chemical Society Reviews, Universal analytical modeling of plasmonic nanoparticles, 2017.
[24] Q. Zhan, Cylindrical vector beams: from mathematical concepts to applications, Advances in Optics and Photonics 1, pp. 1-57, 2009.
[25] S. Mikhailov and K. Ziegler, New electromagnetic mode in graphene, Physical Review Letters, vol. 99, 016803, 2007.
[26] Y. V. Morozov, M. Kuno, Optical constants and dynamic conductivities of single layer MoS2, MoSe2, and WSe2, Appl. Phys. Lett. 107, 083103, 2015.
[27] U. Wurstbauer, B. Miller, E, Parzingerand, A. W Holleitner, Light–matter interaction in transition metal dichalcogenides and their heterostructures, J. Phys. D: Appl. Phys. 50, 173001, 2017.
[28] F. Monticone and A. Alu, Leaky-Wave Theory, Techniques, and Applications: From Microwaves to Visible Frequencies, Proc. IEEE, 103, pp. 793–821, 2015.
[29] C. Forestiere, G. Miano, M. Pascale, R. Tricarico, Quasi-1D Electromagnetic Resonators, arXiv:1806.11088v2, 2018.
[30] P. Debye, Der lichtdruck auf kugeln von beliebigem material, Annalen der physik, vol. 335, pp. 57-136, 1909.
[31] J. A. Stratton, Electromagnetic theory, John Wiley & Sons, 2007.
[32] G. Miano and F. Villone, "A surface integral formulation of maxwell equations for topologically complex conducting domains," IEEE transactions on antennas and propagation, vol. 53, no. 12, pp. 4001–4014, 2005.
[33] C. Forestiere, G. Iadarola, G. Rubinacci, A. Tamburrino, L. Dal Negro, G. Miano, Surface integral formulations for the design of plasmonic nanostructures, JOSA A, 2314-2327, 2012.
[34] A. Doicu, T. Wriedt, and Y. Eremin, Light Scattering by Systems of Particles, Springer-Verlag, 2006.